CASE STUDIES

# A study on the dynamic model of a three-phase grid-connected inverter and an innovative method for its verification

**Arash Esfandiari[1] · Mohammad Naser Hashemnia[1]**

**Abstract** The ever-increasing use of renewable energy sources has underlined the role of power electronic converters as an interface between these resources and the power grid. One application of these converters is in three-phase inverters utilized in a solar power plant to inject active/reactive power to the grid. The dynamic model of power electronic converters is necessary for investigating the overall system stability and the design of the controller for the converters. Generally, the inverter dynamic model is needed to investigate the dynamic behavior of inverters in different applications. This paper is a study of the dynamical model of the grid-connected voltage source inverter, which is extracted by the state-space averaging (SSA) method. This model is verified by applying the values of the operating point to the inverter in Matlab® Simulink environment. To attain the steady-state operating point, the zero component of the duty ratio of the converter is required. To obtain this component, the matrix form of the converter's average equations is used. Overally, using the above methods provides a more efficient and clear understanding of the dynamic model of the converter.

✉ Mohammad Naser Hashemnia
hashemnia@mshdiau.ac.ir

Arash Esfandiari
arash_s11@mshdiau.ac.ir

[1] Faculty of Engineering, Islamic Azad University of Mashhad, Mashhad, Iran

## 1 Introduction

The dynamic model of power electronics converters is necessary to study the overall system stability and design of the controller for the converter and systems based on the converter (Kaviani 2012;

Puukko 2012; Puukko et al. 2012; Sallinen et al. 2018). Grid-connected inverters are the basic components that transmit the power from solar panels to the grid (Wen et al. 2015; Blaabjerg et al. 2; Rocabert et al. 2012).In general, it is necessary to have the dynamic model of the inverter for evaluating the dynamic behavior of inverters in different applications (Puukko et al. 2012; Liu 2017). The task of the control system is to set up a suitable voltage at the converter output and to mitigate interruptions and harmonics produced by the converter to inject the produced electrical energy into the power grid. This shows the importance of attaining a small-signal model of the converter (Middlebrook et al. 1976).

In this paper, we first find the operating point of a given voltage source converter in an analytical form. Then, using time-domain simulations, the operating point of the converter is obtained, similar to that obtained in the analytical form. Indeed, simulation in the time domain was used to confirm the operating point obtained by the circuit analysis.

For this purpose, applying the duty ratio obtained from the linearized average model (i.e., duty ratio in the steady-state) to the converter in simulation, we compared the operating point obtained from the simulation with the operating point extracted analytically. Obtaining the zero component of the duty ratio for this purpose is one innovation of this paper compared to previous papers. The Table 1 gives the parameters of the converter under study.

**Table 1** The parameter of the converter

| Parameter | Value of parameter |
|---|---|
| $f_{SW}$(switching frequency) | 100KHz |
| $U_{oq}$(q channel output voltage) | 0V |
| $U_{od}$(d channel output voltage) | 8.6V |
| $I_{in}$(input current) | 2A |
| $U_{in}$(input voltage) | 30V |
| f (grid frequency) | 50Hz |
| $L_{a,b,c}$(output inductance) | 73$\mu$H |
| $R_L$(resistance of the output inductor) | 15mΩ |
| $R_{on}$(switch on-state resistance) | 0.1Ω |
| $R_S$(Equivalent grid resistance) | 50mΩ |
| $R_{eq} = R_L + R_{on} + R_S$ | 0.165Ω |

Figure 1 presents a grid-connected voltage-fed VSI. The power stage consists of a switch matrix and output inductors. In the case of the VF-VSI in Fig. 1, the system inputs are the input DC voltage ($u_{in}$) and grid phase voltage ($u_{(a,b,c)n}$). On the other hand, the outputs are the input current ($i_{in}$) and grid phase currents ($i_{o(a,b,c)}$) (Puukko 2012). The reason for this is that the solar panel voltage is considered as a constant value on the input side or DC side of the converter. For the output side or the AC side of the converter, the grid voltage connected to the converter is considered as constant AC.

## 2 Average model

SSA modeling begins by determining the voltage and current waveforms of the inductor and capacitor (Erickson and Dragan 2007). Then, the time average valued equations are computed by the time derivatives of state variables $\frac{dx(t)}{d_t}$ and the controllable output variables $y(t)$ (Puukko 2012). In this converter, the state variable $x(t)$ is the output inductor current. The average equations of output voltage inductors are as follows:

$$\langle u_{La}\rangle = \langle u_{AN}\rangle - r_{La}\langle i_{La}\rangle - \langle u_{an}\rangle - \langle u_{nN}\rangle \tag{1}$$

$$\langle u_{Lb}\rangle = \langle u_{BN}\rangle - r_{Lb}\langle i_{Lb}\rangle - \langle u_{bn}\rangle - \langle u_{nN}\rangle \tag{2}$$

$$\langle u_{Lc}\rangle = \langle u_{CN}\rangle - r_{Lc}\langle i_{Lc}\rangle - \langle u_{cn}\rangle - \langle u_{nN}\rangle \tag{3}$$

The angle brackets in the above relationships denote average values. To simplify these equations, $r_L = r_{L(a,b,c)}$ and $r_{eq} = r_{sw} + r_L$ are used. Equivalent series resistance includes the switch on-state resistance $r_{sw}$ and inductor resistance $r_L$. Also, $d_{A,B,C}$ is the duty ratio of the upper switch in the corresponding phase leg (Puukko 2012). (Fig. 2)

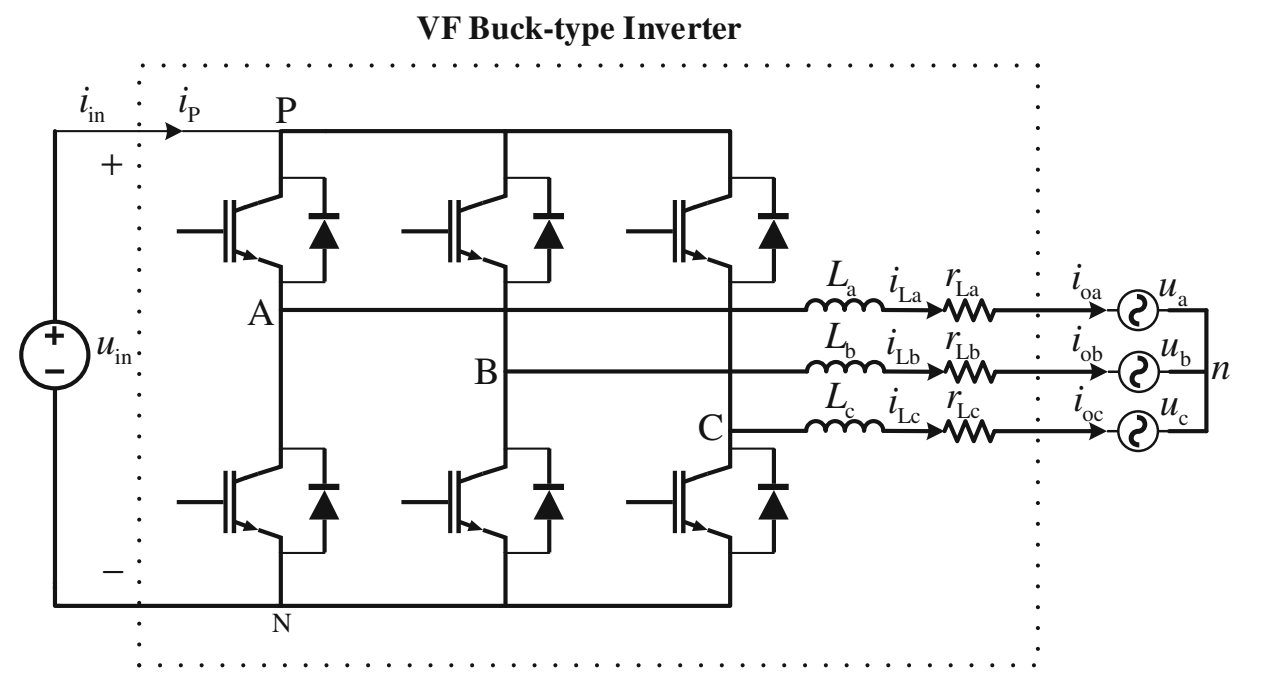

**Fig. 1** A grid-connected three-phase voltage-fed VSI-type inverter

$$\langle u_{La}\rangle = d_A\langle u_{in}\rangle - r_{eq}\langle i_{La}\rangle - \langle u_{an}\rangle - \langle u_{nN}\rangle \tag{4}$$

$$\langle u_{Lb}\rangle = d_B\langle u_{in}\rangle - r_{eq}\langle i_{Lb}\rangle - \langle u_{bn}\rangle - \langle u_{nN}\rangle \tag{5}$$

$$\langle u_{Lc}\rangle = d_C\langle u_{in}\rangle - r_{eq}\langle i_{Lc}\rangle - \langle u_{cn}\rangle - \langle u_{nN}\rangle \tag{6}$$

$$\langle i_{in}\rangle = d_A\langle i_{La}\rangle + d_B\langle i_{Lb}\rangle + d_C\langle i_{Lc}\rangle \tag{7}$$

$$\langle i_{oa}\rangle = \langle i_{La}\rangle \tag{8}$$

$$\langle i_{ob}\rangle = \langle i_{Lb}\rangle \tag{9}$$

$$\langle i_{oc}\rangle = \langle i_{Lc}\rangle \tag{10}$$

According to space vector theory, we can transform a three-phase variable $x_{a,b,c}(t)$ to a single complex value $x(t)$ and a real-valued zero sequence component $x_z(t)$ in the stationary reference frame (Puukko 2012). It is noteworthy that zero components of grid voltage and current under symmetrical conditions are assumed zero in this paper. However, for the duty ratio of the converter, such an assumption is not valid because the duty ratio has a zero-sequence component. This issue is discussed in more detail in the following sections. The real and imaginary components of the space vector in the stationary reference frame are assigned alpha ($x_\alpha$) and beta ($x_\beta$), respectively, which are obtained using the Clark's transformation. This transformation maps the natural abc coordinate system to the stationary coordinate system.

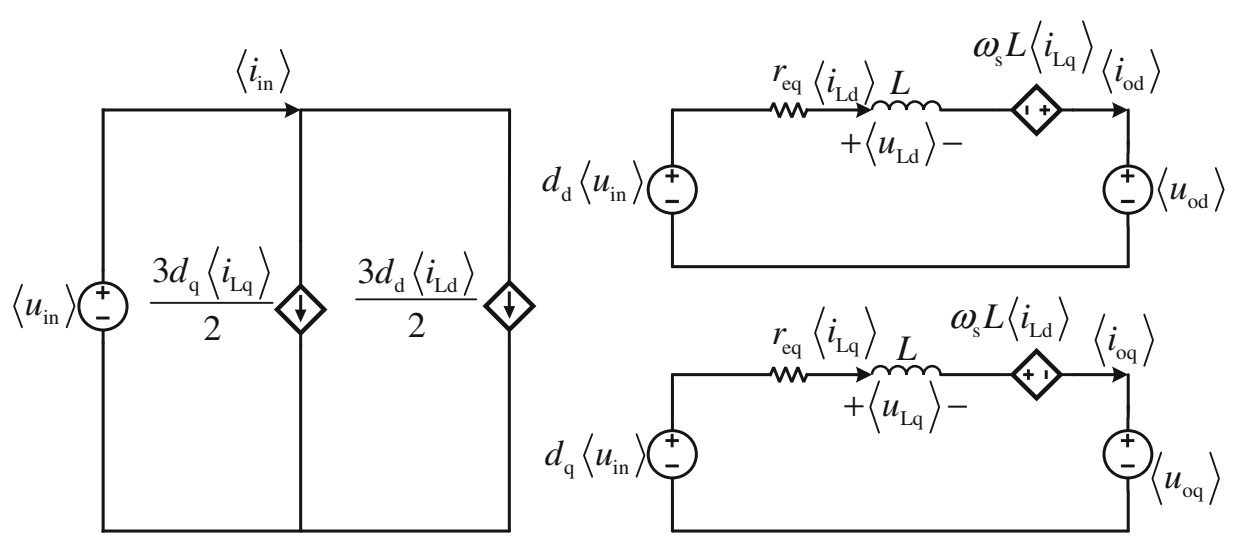

**Fig. 2** The large-signal model of grid-connected three-phase voltage-fed VSI-type inverter (Suntio et al. 2017)

$$x(t) = \frac{2}{3}\left(x_a(t)e^{j0} + x_b(t)e^{j2\pi/3} + x_c(t)e^{j4\pi/3}\right) = |x(t)|e^{j\varphi} = x_\alpha(t) + jx_\beta(t) \tag{11}$$

$$x_z(t) = \frac{1}{3}(x_a(t) + x_b(t) + x_c(t)) \tag{12}$$

The fact that the moduli of the space vector differs from the amplitude of the sinusoidal variable has prompted researchers to introduce some methods to "correct" this supposed deficiency. If the transformation is multiplied by 2/3, a scale change is made in moving from *a-b-c* to α-β-0 variables to eliminate this difference (Holmes and Lipo 2003). With this coefficient, the magnitude of the space vector becomes equal to the peak value of the variables in the symmetrical and balanced three-phase system. This is known as the amplitude invariant form of transformation. If a coefficient of $\sqrt{2/3}$ is used instead, it is called power invariant transformation (Puukko 2012).

Multiplying (4) with $\frac{2}{3}e^{j0}$, (5) with $\frac{2}{3}e^{j2\pi/3}$, and (6) with $\frac{2}{3}e^{j4\pi/3}$ and then summing them together and using (11) yields (Puukko 2012).

$$\langle \boldsymbol{u}_L \rangle = -r_{eq}\langle \boldsymbol{i}_L \rangle + \boldsymbol{d}\langle u_{in} \rangle - \langle \boldsymbol{u}_o \rangle - \frac{2}{3}\overbrace{\left(e^{j0} + e^{j2\pi/3} + e^{j4\pi/3}\right)}^{=0}\langle u_{nN} \rangle = -r_{eq}\langle \boldsymbol{i}_L \rangle + \boldsymbol{d}\langle u_{in} \rangle - \langle \boldsymbol{u}_o \rangle \tag{13}$$

Rewriting (13) results in (14) which yields the relation between voltage and current space vectors. In the following, it is considered that $L = L_{a,b,c}$.

$$\frac{d\langle \boldsymbol{i}_L \rangle}{dt} = \frac{1}{L}\left[-r_{eq}\langle \boldsymbol{i}_L \rangle + \boldsymbol{d}\langle u_{in} \rangle - \langle \boldsymbol{u}_o \rangle\right] \tag{14}$$

where

$$d = \frac{2}{3}\left(d_A e^{j0} + d_B e^{j\frac{2\pi}{3}} + d_C e^{j\frac{4\pi}{3}}\right)$$
$$\langle i_L \rangle = \frac{2}{3}\left(\langle i_{La} \rangle e^{j0} + \langle i_{Lb} \rangle e^{j\frac{2\pi}{3}} + \langle i_{Lb} \rangle e^{j\frac{4\pi}{3}}\right) \tag{15}$$
$$\langle u_o \rangle = \frac{2}{3}\left(\langle u_{an} \rangle e^{j0} + \langle u_{bn} \rangle e^{j\frac{2\pi}{3}} + \langle u_{cn} \rangle e^{j\frac{4\pi}{3}}\right)$$

After transforming the three-phase variables to a stationary (α-β) reference frame, these space vectors are transformed to the synchronous (d-q) reference frame using Park's transformation (Park 1929). Transforming AC electrical quantities into constant DC quantities by Park's transformation is beneficial for analyzing the steady-state stability of power electronic converters (PECs) (Alskran 2014). To apply the SSA method to the inverter, we need to have constant values under steady-state conditions. It is clear that the AC values do not have constant values even at steady-state (Puukko et al. 2012).

$$x^s(t) = x(t)e^{-j\omega_s t} = x_d + jx_q \tag{16}$$

In (16), $\omega_s$ is the grid frequency in rad/s and superscript 's' denotes the synchronous reference frame. The d-q subscripts represent the real and imaginary components of a space-vector in a synchronous-reference-frame, respectively. Transforming both sides of (14) into the synchronous reference frame using (16) yields (Puukko 2012):

$$\frac{d\left(\langle i_L^s \rangle e^{j\omega_s t}\right)}{dt} = \frac{1}{L}d^s\langle u_{in} \rangle e^{j\omega_s t} - \frac{r_{eq}}{L}\langle i_L^s \rangle e^{j\omega_s t} - \frac{1}{L}\langle u_o^s \rangle e^{j\omega_s t}$$
$$\langle i_L^s \rangle j\omega_s e^{j\omega_s t} + \frac{d\langle i_L^s \rangle e^{j\omega_s t}}{dt} = \frac{d^s}{L}\langle u_{in} \rangle e^{j\omega_s t} - \frac{r_{eq}}{L}\langle i_L^s \rangle e^{j\omega_s t} - \frac{1}{L}\langle u_o^s \rangle e^{j\omega_s t}$$
$$e^{j\omega_s t}\left(\langle i_L^s \rangle j\omega_s + \frac{d\langle i_L^s \rangle}{dt}\right) = e^{j\omega_s t}\left(\frac{1}{L}d^s\langle u_{in} \rangle - \frac{r_{eq}}{L}\langle i_L^s \rangle - \frac{1}{L}\langle u_o^s \rangle\right) \tag{17}$$

Finally:

$$\frac{d\langle \boldsymbol{i}_L^s \rangle}{dt} = \frac{1}{L}\left[-(r_{eq} + j\omega_s L)\langle \boldsymbol{i}_L^s \rangle + \boldsymbol{d}^s\langle u_{in} \rangle - \langle \boldsymbol{u}_o^s \rangle\right] \tag{18}$$

The direct and quadrature components of (17) are as follows (Puukko 2012):

$$\frac{d\langle i_{Ld} + ji_{Lq} \rangle}{dt} = \frac{1}{L}\left[-(r_{eq} + j\omega_s L)\langle i_{Ld} + ji_{Lq} \rangle + (d_d + jd_q)\langle u_{in} \rangle - \langle u_{od} + ju_{oq} \rangle\right] \tag{19}$$

$$\frac{d\langle i_{Ld} \rangle}{dt} = \frac{1}{L}\left[-r_{eq}\langle i_{Ld} \rangle + \omega_s L\langle i_{Lq} \rangle + d_d\langle u_{in} \rangle - \langle u_{od} \rangle\right] \tag{20}$$

$$\frac{d\langle i_{Lq} \rangle}{dt} = \frac{1}{L}\left[-\omega_s L\langle i_{Ld} \rangle - r_{eq}\langle i_{Lq} \rangle + d_q\langle u_{in} \rangle - \langle u_{oq} \rangle\right] \tag{21}$$

$$\langle i_{od} \rangle = \langle i_{Ld} \rangle \tag{22}$$

$$\langle i_{oq} \rangle = \langle i_{Lq} \rangle \tag{23}$$

The input current in (7) can be expressed as follows (Puukko 2012):

$$\langle i_{in} \rangle = d_A\langle i_{La} \rangle + d_B\langle i_{Lb} \rangle + d_C\langle i_{Lc} \rangle = \frac{3}{2}\mathrm{Re}\left\{\boldsymbol{d}^s\langle \boldsymbol{i}_L^s \rangle^*\right\} \tag{24}$$

Further simplification yields:

$$\langle i_{in} \rangle = \frac{3}{2}\mathrm{Re}\left\{\boldsymbol{d}^s e^{-j\omega_s t}\left(\langle \boldsymbol{i}_L^s \rangle e^{-j\omega_s t}\right)^*\right\} = \frac{3}{2}\mathrm{Re}\left\{\boldsymbol{d}^s\langle \boldsymbol{i}_L^{s*} \rangle\right\} \tag{25}$$

Using the following relationships:

$$\begin{aligned} \boldsymbol{d}^{\mathrm{s}} &= d_{\mathrm{d}} + \mathrm{j}d_{\mathrm{q}} \\ \boldsymbol{i}_{\mathrm{L}}^{\mathrm{s}} &= i_{\mathrm{Ld}} + \mathrm{j}i_{\mathrm{Lq}} \end{aligned} \tag{26}$$

The switching time averages input current of the inverter is obtained as:

$$\begin{aligned} \langle i_{\mathrm{in}} \rangle &= \frac{3}{2}\mathrm{Re}\left\{ \left(d_{\mathrm{d}} + \mathrm{j}d_{\mathrm{q}}\right)\left\langle i_{\mathrm{Ld}} - \mathrm{j}i_{\mathrm{Lq}} \right\rangle \right\} \\ &= \frac{3}{2}\left[d_{\mathrm{d}}\langle i_{\mathrm{Ld}}\rangle + d_{\mathrm{q}}\langle i_{\mathrm{Lq}}\rangle\right] \end{aligned} \tag{27}$$

Equations (20)–(24) are known as the large-signal model of VF-VSI in the synchronous-reference-frame (Puukko 2012):

## 3 Operating point

To obtain the operating point of VF-VSI in steady-state for the inductor current, capacitor voltage, and duty ratio, it is necessary to have the input voltage, input current, and output voltage. In this way, by putting the derivatives equal to zero and placing the average values with their corresponding steady-state values, we can get the operating point in the steady-state. In other words, due to the sinusoidal and variable nature of inductor currents and capacitor voltages in the output port of inverter, we cannot consider constant steady-state values for these variables. However, as stated in the previous section, by transforming these variables into the synchronous reference frame, it is possible to obtain DC values for these variables. Thus, the average values of inductor voltage and capacitor current in steady-state are DC and their time derivative is zero, which means that:

$$\begin{aligned} \left\langle I_{C_{\mathrm{d,q}}} \right\rangle = 0 \rightarrow \left\langle C\frac{\mathrm{d}u_{C_{\mathrm{d,q}}}}{\mathrm{d}t} \right\rangle = 0 \rightarrow C\left\langle \frac{\mathrm{d}u_{C_{\mathrm{d,q}}}}{\mathrm{d}t} \right\rangle = 0 \\ \left\langle U_{L_{\mathrm{d,q}}} \right\rangle = 0 \rightarrow \left\langle L\frac{\mathrm{d}i_{L_{\mathrm{d,q}}}}{\mathrm{d}t} \right\rangle = 0 \rightarrow L\left\langle \frac{\mathrm{d}i_{L_{\mathrm{d,q}}}}{\mathrm{d}t} \right\rangle = 0 \end{aligned} \tag{28}$$

Consider the following relationships (Puukko 2012):

$$-r_{\mathrm{eq}}I_{\mathrm{Ld}} + \overbrace{\omega_{\mathrm{s}}LI_{\mathrm{Lq}}}^{=0} + D_{\mathrm{d}}U_{\mathrm{in}} - U_{\mathrm{od}} = 0 \tag{29}$$

$$\overbrace{-r_{\mathrm{eq}}I_{\mathrm{Lq}}}^{=0} - \omega_{\mathrm{s}}LI_{\mathrm{Ld}} + D_{\mathrm{q}}U_{\mathrm{in}} - U_{\mathrm{oq}} = 0 \tag{30}$$

$$I_{\mathrm{in}} = \frac{3}{2}\left[D_{\mathrm{d}}I_{\mathrm{Ld}} + D_{\mathrm{q}}I_{\mathrm{Lq}}\right] \tag{31}$$

$$I_{\mathrm{od}} = I_{\mathrm{Ld}} \tag{32}$$

$$I_{\mathrm{oq}} = I_{\mathrm{Lq}} \tag{33}$$

Equations (29) and (30) are the voltage equations in d-channel and q-chanel, respectively. It should be noted that transformer voltage terms have been omitted because synchronous reference frame inductor currents ($I_{\mathrm{Ld}}$ and $I_{\mathrm{Lq}}$) are DC at steady state.

The adopted control strategy is grid voltage oriented vector control, wherein the d-axis is supposed to be aligned with the inverter's output voltage space vector. This grid voltage orientation makes possible decoupled control of active and reactive powers. This is so because the active power will only depend on the d-axis current whereas the reactive power will only depend on its q-axis component.

This orientation results in the following:

$$U_{\mathrm{oq}} = 0 \tag{34}$$

Moreover, it is customary to keep the converter operate at unity power factor; this results in zero value for the steady-state value of grid current's q-channel:

$$I_{\mathrm{Lq}} = 0 \tag{35}$$

Under such conditions, only the real power is injected to the grid via the converter and the power injected is given in terms of space vectors in the synchronous reference frame as follows:

$$\begin{aligned} &\begin{cases} \boldsymbol{u}_{\mathrm{o}} = U_{\mathrm{od}} + \mathrm{j}U_{\mathrm{oq}} \\ \boldsymbol{i}_{\mathrm{L}} = I_{\mathrm{Ld}} + \mathrm{j}I_{\mathrm{Lq}} \end{cases} \\ &\boldsymbol{S} = \langle \boldsymbol{u}_{\mathrm{o}} \rangle \langle \boldsymbol{i}_{\mathrm{L}}^{*} \rangle \end{aligned} \tag{36}$$

By rewriting the above equation, we have:

$$\begin{aligned} \boldsymbol{S} &= \left(U_{\mathrm{od}} + \mathrm{j}U_{\mathrm{oq}}\right)\left(I_{\mathrm{Ld}} - jI_{\mathrm{Lq}}\right) \\ &= \left[U_{\mathrm{od}}I_{\mathrm{Ld}} + U_{\mathrm{oq}}I_{\mathrm{Lq}} + \mathrm{j}\left(U_{\mathrm{oq}}I_{\mathrm{Ld}} - U_{\mathrm{od}}I_{\mathrm{Lq}}\right)\right] \\ \boldsymbol{P} &= U_{\mathrm{od}}I_{\mathrm{Ld}} \end{aligned} \tag{37}$$

In (37), only real power is injected. In the following, the steady-state value for d-channel inductor current $I_{\mathrm{Ld}}$ is obtained by substituting (35) in (31) (Puukko 2012):

$$I_{\mathrm{Ld}} = \frac{2}{3}\frac{I_{\mathrm{in}}}{D_{\mathrm{d}}} \tag{38}$$

By substituting (35) and (38) in (29) the quadratic equation for the d-channel duty ratio ($D_{\mathrm{d}}$) is obtained (Puukko 2012):

$$U_{\mathrm{in}}D_{\mathrm{d}}^2 - U_{\mathrm{od}}D_{\mathrm{d}} - \frac{2}{3}r_{\mathrm{eq}}I_{\mathrm{in}} = 0 \tag{39}$$

which can be solved as:

$$D_{\mathrm{d}} = \frac{U_{\mathrm{od}} + \sqrt{U_{\mathrm{od}}^2 + \frac{8}{3}r_{\mathrm{eq}}U_{\mathrm{in}}I_{\mathrm{in}}}}{2U_{\mathrm{in}}} \tag{40}$$

By substituting (40) in (38) and rewriting the new equation, d-channel inductor current ($I_{\mathrm{Ld}}$) can be calculated as follows:

$$I_{\mathrm{Ld}} = \frac{4}{3}\frac{U_{\mathrm{in}}I_{\mathrm{in}}}{U_{\mathrm{od}} + \sqrt{U_{\mathrm{od}}^2 + \frac{8}{3}r_{\mathrm{eq}}U_{\mathrm{in}}I_{\mathrm{in}}}} \tag{41}$$

The steady-state value for q-channel duty ratio ($D_{\mathrm{q}}$) can be obtained by substituting (34) and (35) in (30) (Puukko 2012):

$$D_{\mathrm{q}} = \frac{2\omega_{\mathrm{s}}LI_{\mathrm{in}}}{3U_{\mathrm{in}}D_{\mathrm{d}}} = \frac{4}{3}\frac{\omega_{\mathrm{s}}LI_{\mathrm{in}}}{U_{\mathrm{od}} + \sqrt{U_{\mathrm{od}}^2 + \frac{8}{3}r_{\mathrm{eq}}U_{\mathrm{in}}I_{\mathrm{in}}}} \tag{42}$$

Equations (34), (35), and (40)–(42) and the predefined values $U_{\mathrm{in}}$, $I_{\mathrm{in}}$, and $U_{\mathrm{od}}$ are defined as the operating point of a VF-VSI (Puukko 2012). The input voltage and current of the inverter are the predefined values obtained by the linearization of the converter at those points. The output voltage of the inverter is also as same as the grid voltage connected to the output port of VF-VSI. Therefore, these values are considered as predefined values. Table 2 presents the operating point of inverter obtained using the data in Table 1 and (40)–(42):

## 4 Linearization of transformed average model

Large-signal average model equations of VF-VSI have a set of nonlinear terms. A linearized model facilitates the analysis of the model. To build a linear model, small-signal equations are derived around a predefined operating point. It is noteworthy that most of the analysis and AC modeling methods, such as Laplace transform and other frequency-domain methods, are not applicable to nonlinear systems. So, for analyzing these circuits and correct prediction of poles and zeros of small-signal transfer functions of switching converter, the nonlinear large-signal equations must be linearized. To construct a small-signal model around a predefined or quiescent operating point, it is assumed that the dq voltages and currents and duty cycle of the converter are equal to some given quiescent or DC values, plus some superimposed small AC variations or perturbations (Erickson and Dragan and Dragan 2007). Hence, we have:

**Table 2** The operating point of voltage fed inverter

| The operating point of voltage fed inverter |
|---|
| $I_{\mathrm{oq}} = 0\mathrm{A}$ |
| $I_{\mathrm{od}} = 4.4236$ |
| $D_{\mathrm{d}} = 0.3103$ |
| $D_{\mathrm{q}} = 0.0033$ |

$$\begin{aligned}
\langle i_{\mathrm{Ld}}\rangle_{T_{\mathrm{s}}} &= I_{\mathrm{Ld}} + \hat{i}_{\mathrm{Ld}}(t)\\
\langle i_{\mathrm{Lq}}\rangle_{T_{\mathrm{s}}} &= I_{\mathrm{Lq}} + \hat{i}_{\mathrm{Lq}}(t)\\
\langle u_{\mathrm{od}}\rangle_{T_{\mathrm{s}}} &= U_{\mathrm{od}} + \hat{u}_{\mathrm{od}}(t)\\
\langle u_{\mathrm{oq}}\rangle_{T_{\mathrm{s}}} &= U_{\mathrm{oq}} + \hat{u}_{\mathrm{oq}}(t)\\
\langle u_{\mathrm{in}}\rangle_{T_{\mathrm{s}}} &= U_{\mathrm{in}} + \hat{u}_{\mathrm{in}}(t)\\
\langle i_{\mathrm{in}}\rangle_{T_{\mathrm{s}}} &= I_{\mathrm{in}} + \hat{i}_{\mathrm{in}}(t)\\
d(t) &= D + \hat{d}(t)
\end{aligned} \tag{43}$$

It is assumed that the AC variations are relatively smaller compared to the DC quiescent values, i.e. (Erickson and Dragan 2007):

$$\begin{aligned}
|I_{Ld}| &\gg |\hat{i}_{Ld}(t)|\\
|I_{Lq}| &\gg |\hat{i}_{Lq}(t)|\\
|U_{od}| &\gg |\hat{u}_{od}(t)|\\
|U_{oq}| &\gg |\hat{u}_{oq}(t)|\\
|U_{in}| &\gg |\hat{u}_{in}(t)|\\
|I_{in}| &\gg |\hat{i}_{in}(t)|\\
|D| &\gg |\hat{d}(t)|
\end{aligned} \tag{44}$$

Accordingly, the nonlinear Eqs. (20)–(24) can be linearized by inserting the relevant terms from Eq. (44) (Erickson and Dragan 2007). For instance, Eq. (20) or d-channel equation of inverter is given as follows:

$$\begin{aligned}
\frac{\mathrm{d}\langle I_{\mathrm{Ld}}+\hat{i}_{\mathrm{Ld}}\rangle}{\mathrm{d}t} &= \frac{1}{L}\left[-r_{\mathrm{eq}}\langle I_{\mathrm{Ld}}+\hat{i}_{\mathrm{Ld}}\rangle + \omega_{\mathrm{s}}L\langle I_{\mathrm{Lq}}+\hat{i}_{\mathrm{Lq}}\rangle + (D_{\mathrm{d}}+\hat{d}_{\mathrm{d}})\langle U_{\mathrm{in}}+\hat{u}_{\mathrm{in}}\rangle - \langle U_{\mathrm{od}}+\hat{u}_{\mathrm{od}}\rangle\right]\\
&= \frac{1}{L}\left[\underbrace{\left(-r_{\mathrm{eq}}I_{\mathrm{Ld}} + \omega_{\mathrm{s}}LI_{\mathrm{Lq}} + D_{\mathrm{d}}U_{\mathrm{in}} - U_{\mathrm{od}}\right)}_{\text{Dcterms}} + \underbrace{\left(\hat{d}_{\mathrm{d}}\hat{u}_{\mathrm{in}}\right)}_{\substack{2^{nd}\text{orderacterms}\\ \text{(nonlinear)}}} + \underbrace{\left(-r_{\mathrm{eq}}\hat{i}_{\mathrm{Ld}} + \omega_{\mathrm{s}}L\hat{i}_{\mathrm{Lq}} + D_{\mathrm{d}}\hat{u}_{\mathrm{in}} + \hat{d}_{\mathrm{d}}U_{\mathrm{in}} - \hat{u}_{\mathrm{od}}\right)}_{1^{st}\text{orderacterms(linear)}}\right]
\end{aligned} \tag{45}$$

The derivative of DC (constant) terms is zero. To construct a small-signal AC model of the inverter, the DC terms can be considered known constant quantities. The first-order AC terms contain a single AC quantity, usually multiplied by a constant coefficient or a DC term. These terms are the linear functions of the AC perturbation. The second-order AC terms contain the products of two AC quantities. Therefore, they are nonlinear because they have the multiplication of time-varying signals. So it is desired to neglect the nonlinear AC terms. In fact, each of the second-order nonlinear terms is much smaller in magnitude than one or more linear first-order AC terms. Also, the DC terms on the right-hand side of the Eq. (45) are equal to the DC terms on the left-hand side (Erickson and Dragan 2007). Finally, Eq. (45) is left with the first-order AC terms on both sides of the equation. Hence:

$$\frac{\mathrm{d}\hat{i}_{\mathrm{Ld}}}{\mathrm{d}t}=\frac{1}{L}\left[-r_{\mathrm{eq}}\hat{i}_{\mathrm{Ld}}+\omega_{\mathrm{s}}L\hat{i}_{\mathrm{Lq}}+D_{\mathrm{d}}\hat{u}_{\mathrm{in}}+\hat{d}_{\mathrm{d}}U_{\mathrm{in}}-\hat{u}_{\mathrm{od}}\right] \quad (46)$$

The remaining small-signal equations are as follows (Puukko 2012):

$$\frac{\mathrm{d}\hat{i}_{\mathrm{Lq}}}{\mathrm{d}t}=\frac{1}{L}\left[-r_{\mathrm{eq}}\hat{i}_{\mathrm{Lq}}+\omega_{\mathrm{s}}L\hat{i}_{\mathrm{Ld}}+D_{\mathrm{q}}\hat{u}_{\mathrm{in}}+\hat{d}_{\mathrm{q}}U_{\mathrm{in}}-\hat{u}_{\mathrm{oq}}\right] \quad (47)$$

$$\hat{i}_{\mathrm{in}}=\frac{3}{2}\left[D_{\mathrm{d}}\hat{i}_{\mathrm{Ld}}+D_{\mathrm{q}}\hat{i}_{\mathrm{Lq}}+I_{\mathrm{Ld}}\hat{d}_{\mathrm{d}}+I_{\mathrm{Lq}}\hat{d}_{\mathrm{q}}\right] \quad (48)$$

Equation (48) can be simplified using (35) and (38) (Puukko 2012):

$$\hat{i}_{\mathrm{in}}=\frac{3}{2}\left[D_{\mathrm{d}}\hat{i}_{\mathrm{Ld}}+D_{\mathrm{q}}\hat{i}_{\mathrm{Lq}}+\frac{2}{3}\frac{I_{\mathrm{in}}}{D_{\mathrm{d}}}\hat{d}_{\mathrm{d}}\right] \quad (49)$$

And:

$$\hat{i}_{\mathrm{od}}=\hat{i}_{\mathrm{Ld}} \quad (50)$$

$$\hat{i}_{\mathrm{oq}}=\hat{i}_{\mathrm{Lq}} \quad (51)$$

Equations (46)–(51) denote the VF-VSI small-signal equations around a steady state operating point.

## 5 State space averaging (SSA)

The linearized equations can be rewritten in matrix form using the expression and description of the state space equations. Using this method, the small-signal model of the PWM switching converter is obtained. One of the benefits of this method is the generalization of the results of this modeling method because by applying state-space averaging equations, it is always possible to obtain a small-signal model of converter (Erickson and Dragan 2007). This method, as one of the most well-known and common methods of modeling switching converters, was developed by Middlebrook in the 1970s. It has been extensively studied in (Middlebrook et al. 1976; Puukko 2012; Wester et al. 18). The average-valued equations are as presented in (52), where the angle brackets denote average values and bold-italic fonts define vectors (Puukko 2012):

$$\begin{aligned}\frac{\mathrm{d}\langle \boldsymbol{x}(t)\rangle}{\mathrm{d}t}&=\boldsymbol{f}_1(\langle \boldsymbol{x}(t)\rangle,\langle \boldsymbol{u}(t)\rangle)\\ \langle \boldsymbol{y}(t)\rangle&=\boldsymbol{f}_2(\langle \boldsymbol{x}(t)\rangle,\langle \boldsymbol{u}(t)\rangle)\end{aligned} \quad (52)$$

The state-space averaging method deals with three types of variables involved in modeling the dynamic behavior of the converter. The first of them is the state variable ($\boldsymbol{x}(t)$), which indicates the inductor currents and capacitor voltages in the circuit. Based on dynamic systems state-space theory, there is no limit to selecting the type of state variables. This freedom to choose the state variable is one of the advantages of the state space method (Ogata and Yanjuan 10). However, due to the simplicity to measure the values of inductor currents and capacitor voltages, they are considered as state variables. The next parameters are input or uncontrollable (Puukko 2012) variables $\boldsymbol{u}(t)$. In this study, the input of the converter is connected to a constant voltage source such as a solar panel. The output voltage of the converter is uncontrollable due to the connection to the grid, and hence, when the two-port representation of this system is accepted, independent or uncontrollable input and output voltages at both ends of the converter are considered as input variables $\boldsymbol{u}(t)$ of the SSA-model (Desoer and Ernest 3).

Also, output variables of the system, denoted by $\boldsymbol{y}(t)$, are taken to be dependent variables, which are the current injected to the grid and input current from the solar panel to the converter. So,

$$\begin{aligned}\hat{}&=\begin{bmatrix}\hat{i}_{\mathrm{Ld}} & \hat{i}_{\mathrm{Lq}}\end{bmatrix}^T\\ \hat{}&=\begin{bmatrix}\hat{u}_{\mathrm{in}} & \hat{u}_{\mathrm{od}} & \hat{u}_{\mathrm{oq}} & \hat{d}_{\mathrm{d}} & \hat{d}_{\mathrm{q}}\end{bmatrix}^T\\ \hat{}&=\begin{bmatrix}\hat{i}_{\mathrm{in}} & \hat{i}_{\mathrm{od}} & \hat{i}_{\mathrm{oq}}\end{bmatrix}^T\end{aligned} \quad (53)$$

$$\begin{aligned}\frac{\mathrm{d}\hat{\boldsymbol{x}}(t)}{\mathrm{d}t}&=\boldsymbol{A}\hat{\boldsymbol{x}}(t)+\boldsymbol{B}\hat{\boldsymbol{u}}(t)\\ \hat{\boldsymbol{y}}(t)&=\boldsymbol{C}\hat{\boldsymbol{x}}(t)+\boldsymbol{D}\hat{\boldsymbol{u}}(t)\end{aligned} \quad (54)$$

Equation (54) presents the state space equations in the time domain, which can be transformed to the frequency domain using Laplace transformation to obtain the converter transfer function,

$$\begin{aligned}\frac{\mathrm{d}\hat{\boldsymbol{x}}(t)}{\mathrm{d}t}&=\boldsymbol{A}\hat{\boldsymbol{x}}(t)+\boldsymbol{B}\hat{\boldsymbol{u}}(t)\rightarrow s\boldsymbol{X}(s)=\boldsymbol{A}\boldsymbol{X}(s)+\boldsymbol{B}\boldsymbol{U}(s)\\ \hat{\boldsymbol{y}}(t)&=\boldsymbol{C}\hat{\boldsymbol{x}}(t)+\boldsymbol{D}\hat{\boldsymbol{u}}(t)\rightarrow \boldsymbol{Y}(s)=\boldsymbol{C}\boldsymbol{X}(s)+\boldsymbol{D}\boldsymbol{U}(s)\end{aligned} \quad (55)$$

The state matrices $\boldsymbol{A}$, $\boldsymbol{B}$, $\boldsymbol{C}$ and $\boldsymbol{D}$ can be obtained using the linearized converter Eqs. (46)–(51) and the state space representation of (55) (Puukko 2012):

$$\mathbf{A}=\begin{bmatrix}-\frac{r_{\mathrm{eq}}}{L} & \omega_{\mathrm{s}}\\ -\omega_{\mathrm{s}} & -\frac{r_{\mathrm{eq}}}{L}\end{bmatrix},\mathbf{B}=\begin{bmatrix}\frac{D_{\mathrm{d}}}{L} & -\frac{1}{L} & 0 & \frac{U_{\mathrm{in}}}{L} & 0\\ \frac{D_{\mathrm{q}}}{L} & 0 & -\frac{1}{L} & 0 & \frac{U_{\mathrm{in}}}{L}\end{bmatrix}$$
$$\mathbf{C}=\begin{bmatrix}\frac{3}{2}D_{\mathrm{d}} & \frac{3}{2}D_{\mathrm{q}}\\ 1 & 0\\ 0 & 1\end{bmatrix},\mathbf{D}=\begin{bmatrix}0 & 0 & 0 & \frac{I_{\mathrm{in}}}{D_{\mathrm{d}}} & 0\\ 0 & 0 & 0 & 0 & 0\\ 0 & 0 & 0 & 0 & 0\end{bmatrix} \quad (56)$$

The state matrices in (56) are used to obtain the transfer function from the inputs to the outputs (Puukko 2012):

$$\boldsymbol{Y}(s)=\left[C(s\mathbf{I}-\mathbf{A})^{-1}\ \mathbf{B}+\mathbf{D}\right]\boldsymbol{U}(s)=\mathbf{G}_{\mathrm{Y}}\ \boldsymbol{U}(s) \quad (57)$$

Matrix $\boldsymbol{G}_{\mathrm{Y}}$ is known as the VF-VSI transfer function matrix. The subscript Y indicates that the transfer function matrix is of admittance type. That means that the converter is supplied by two voltages as the inputs and resulting currents are taken as the outputs. Actually, these voltage sources are uncontrollable and can be presented as inputs as seen in (53) and in the state-space Eq. (57). Moreover, the output and input currents of the VF-VSI are dependent on the voltages and are therefore controllable variables. Hence, they can be considered as outputs of the converter.

To clarify this fact in more detail, the reason that output voltage is considered one of the control inputs of the converter is that the value of the output voltage is determined by the grid to which the converter is connected to and we do not have control over it. Same as the grid voltage, there is no control over the converter's input voltage coming from battery or solar panels (if modeled as voltage sources). So they are converter inputs in the SSA model. On the other hand, the output current of the converter being injected to the grid is controllable by the switching pattern of the converter, hence being considered as a control output. A similar reasoning justifies consideration of the input current to the converter as one of the control inputs. It is also clear that the duty ratio is another control input of the system under consideration.

Matrix $\boldsymbol{G}_{\mathrm{Y}}$ is as follows (Puukko 2012):

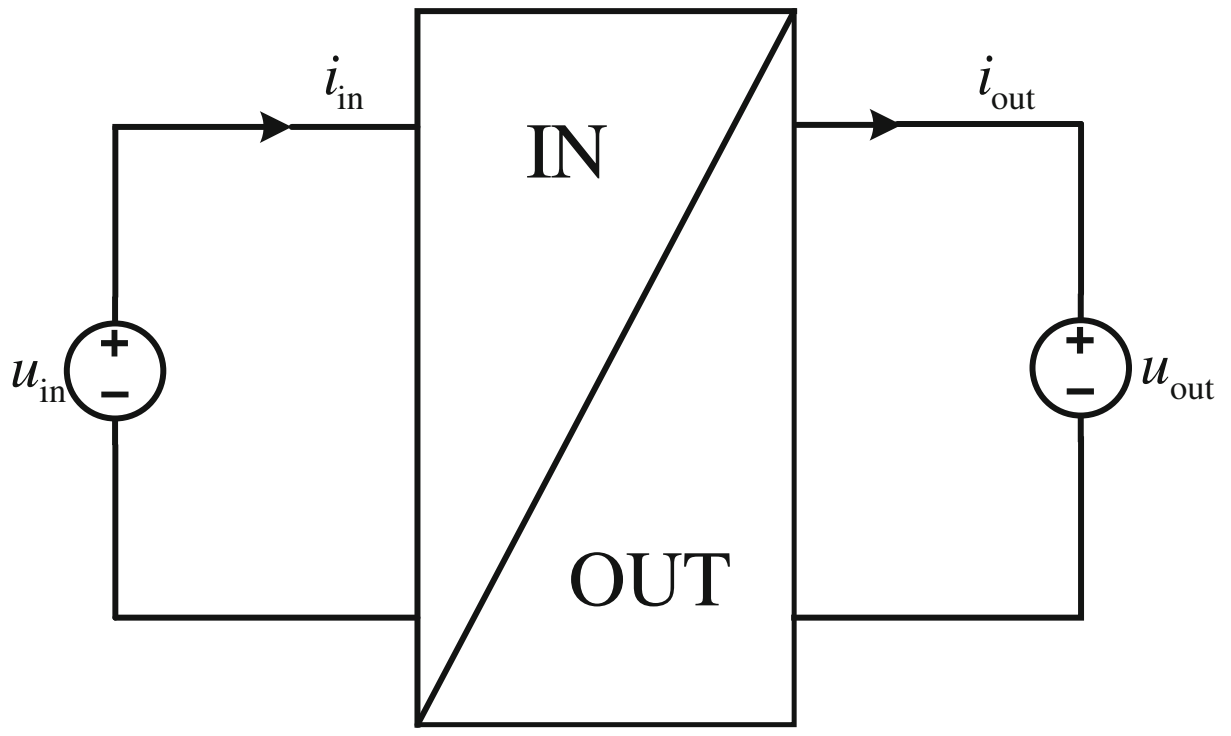

**Fig. 3** A voltage-to-voltage converter (Messo 8)

$$\begin{bmatrix} \hat{i}_{\mathrm{in}} \\ \hat{i}_{\mathrm{od}} \\ \hat{i}_{\mathrm{oq}} \end{bmatrix} = \begin{bmatrix} Y^{\mathrm{Y}}_{\mathrm{in}} & T^{\mathrm{Y}}_{\mathrm{oi-d}} & T^{\mathrm{Y}}_{\mathrm{oi-q}} & G^{\mathrm{Y}}_{\mathrm{ci-d}} & G^{\mathrm{Y}}_{\mathrm{ci-q}} \\ G^{\mathrm{Y}}_{\mathrm{io-d}} & -Y^{\mathrm{Y}}_{\mathrm{o-d}} & G^{\mathrm{Y}}_{\mathrm{cr-qd}} & G^{\mathrm{Y}}_{\mathrm{co-d}} & G^{\mathrm{Y}}_{\mathrm{co-qd}} \\ G^{\mathrm{Y}}_{\mathrm{io-q}} & G^{\mathrm{Y}}_{\mathrm{cr-dq}} & -Y^{\mathrm{Y}}_{\mathrm{o-q}} & G^{\mathrm{Y}}_{\mathrm{co-dq}} & G^{\mathrm{Y}}_{\mathrm{co-q}} \end{bmatrix} \begin{bmatrix} \hat{u}_{\mathrm{in}} \\ \hat{u}_{\mathrm{od}} \\ \hat{u}_{\mathrm{oq}} \\ \hat{d}_{\mathrm{d}} \\ \hat{d}_{\mathrm{q}} \end{bmatrix} \tag{58}$$

Using the VF-VSI transfer functions in (58), a linear model can be obtained as shown in Fig. 3 (Puukko 2012). (Fig. 4)

In the following, converter transfer function in matrix $\boldsymbol{G}_{\mathrm{Y}}$ neglecting $r_{\mathrm{eq}}$ can be given as follows. The input admittance is given by (59) (Puukko 2012):

$$Y^{\mathrm{Y}}_{\mathrm{in}} = \frac{\hat{i}_{\mathrm{in}}}{\hat{u}_{\mathrm{in}}} = \frac{3}{2}\frac{D_{\mathrm{d}}^2 + D_{\mathrm{q}}^2}{L}\frac{s}{\Delta_{\mathrm{Y}}} \tag{59}$$

The transfer functions from output voltages dq components to the input current are given by (Puukko 2012):

$$T^{\mathrm{Y}}_{\mathrm{oi-d}} = \frac{\hat{i}_{\mathrm{in}}}{\hat{u}_{\mathrm{od}}} = -\frac{3}{2}\frac{D_{\mathrm{d}}}{L}\left(s - \frac{D_{\mathrm{q}}\omega_{\mathrm{s}}}{D_{\mathrm{d}}}\right)\frac{1}{\Delta_{\mathrm{Y}}} \tag{60}$$

$$T^{\mathrm{Y}}_{\mathrm{oi-q}} = \frac{\hat{i}_{\mathrm{in}}}{\hat{u}_{\mathrm{oq}}} = -\frac{3}{2}\frac{D_{\mathrm{q}}}{L}\left(s - \frac{D_{\mathrm{d}}\omega_{\mathrm{s}}}{D_{\mathrm{q}}}\right)\frac{1}{\Delta_{\mathrm{Y}}} \tag{61}$$

The transfer functions from the control variables (duty ratio dq components) to the input current are given by (Puukko 2012):

$$G^{\mathrm{Y}}_{\mathrm{ci-d}} = \frac{\hat{i}_{\mathrm{in}}}{\hat{d}_{\mathrm{d}}} = \frac{I_{\mathrm{in}}}{D_{\mathrm{d}}}\left(s + \frac{3}{2}\frac{D_{\mathrm{d}}^2 U_{\mathrm{in}}}{L I_{\mathrm{in}}}\right)\frac{s}{\Delta_{\mathrm{Y}}} \tag{62}$$

$$G^{\mathrm{Y}}_{\mathrm{ci-q}} = \frac{\hat{i}_{\mathrm{in}}}{\hat{d}_{\mathrm{q}}} = \frac{3}{2}\frac{D_{\mathrm{q}} U_{\mathrm{in}}}{L}\left(s + \frac{D_{\mathrm{q}}\omega_{\mathrm{s}}}{D_{\mathrm{d}}}\right)\frac{1}{\Delta_{\mathrm{Y}}} \tag{63}$$

The transfer functions from the input voltage to the output current dq components are given by (Puukko 2012):

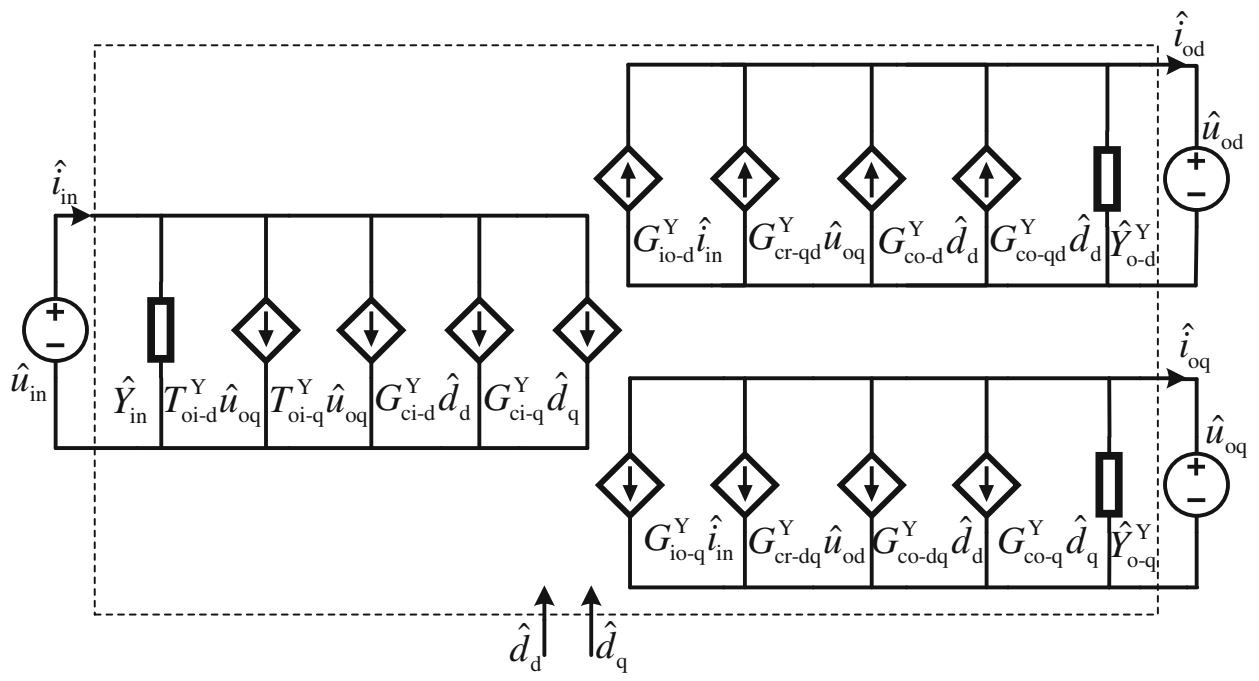

**Fig. 4** The linear model (small-signal) of a grid-connected VF-VSI (Puukko 2012)

$$G^{Y}_{io-d} = \frac{\hat{i}_{od}}{\hat{u}_{in}} = \frac{D_d}{L}\left(s + \frac{D_q \omega_s}{D_d}\right)\frac{1}{\Delta_Y} \quad (64)$$

$$G^{Y}_{io-q} = \frac{\hat{i}_{oq}}{\hat{u}_{in}} = \frac{D_q}{L}\left(s - \frac{D_d \omega_s}{D_q}\right)\frac{1}{\Delta_Y} \quad (65)$$

The d and q channel output admittances and the cross-coupling transfer functions can be given as follows (Puukko 2012):

$$Y^{Y}_{o-d} = -\frac{\hat{i}_{od}}{\hat{u}_{od}} = \frac{s}{L}\frac{1}{\Delta_Y} \quad (66)$$

$$Y^{Y}_{o-q} = -\frac{\hat{i}_{oq}}{\hat{u}_{oq}} = \frac{s}{L}\frac{1}{\Delta_Y} \quad (67)$$

$$G^{Y}_{cr-qd} = \frac{\hat{i}_{od}}{\hat{u}_{oq}} = -\frac{\omega_s}{L}\frac{1}{\Delta_Y} \quad (68)$$

$$G^{Y}_{cr-dq} = \frac{\hat{i}_{oq}}{\hat{u}_{od}} = \frac{\omega_s}{L}\frac{1}{\Delta_Y} \quad (69)$$

The reason for the negative sign of output admittances transfer functions in matrix $\boldsymbol{G}_Y$ is that the output flow direction in a two-port network is opposed to the output current of the small-signal model in this paper. Figure 5. shows a two-port network:

The transfer functions from dq components of duty ratio to dq compnents of output current can be given as follows (Puukko 2012):

$$G^{Y}_{co-d} = \frac{\hat{i}_{od}}{\hat{d}_d} = \frac{U_{in}s}{L}\frac{1}{\Delta_Y} \quad (70)$$

$$G^{Y}_{co-q} = \frac{\hat{i}_{oq}}{\hat{d}_q} = \frac{U_{in}s}{L}\frac{1}{\Delta_Y} \quad (71)$$

$$G^{Y}_{co-qd} = \frac{\hat{i}_{od}}{\hat{d}_q} = \frac{U_{in}\omega_s}{L}\frac{1}{\Delta_Y} \quad (72)$$

$$G^{Y}_{co-dq} = \frac{\hat{i}_{oq}}{\hat{d}_d} = -\frac{U_{in}\omega_s}{L}\frac{1}{\Delta_Y} \quad (73)$$

At the end the determinant $\Delta_Y$ of the transfer functions can be expressed by (74) (Puukko 2012):

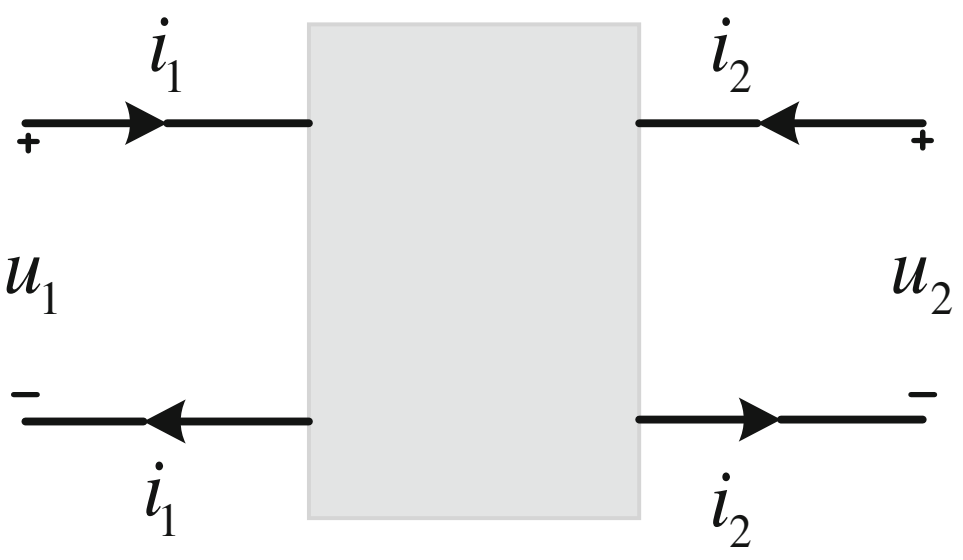

**Fig. 5** A two-port network (Holmes and Lipo 2003)

$$\Delta_Y = s^2 + \omega_s^2 \quad (74)$$

## 6 Simulation setup

In this section, using Matlab® Simulink, a model of VF-VSI is designed to verify the operating points in Table 2. The operating points are obtained in the theoretical form in Sect. 3 using the large-signal model of VF-VSI. The purpose is using time-domain simulations applying a predefined operating point ($U_{in}$, $D_{d,q}$ and $U_{od,oq}$) to obtain the current of the operating point in Sect. 3 ($I_{in}$ and $I_{od,oq}$). By applying the input variables of VF-VSI, it is expected that the output variables of VF-VSI be identical to the analytical calculations (Fig. 6).

Figure 7 shows how to apply the grid output voltage to the output port of the converter.

Figure 8 presents how to produce the d-q channel of duty ratio by inverse Park's transformation.

Firstly, duty ratios of the operating point are applied to the converter without considering the zero component of the duty ratio (i.e. $D_0 = 0$). Figure 9 illustrates time variation the duty ratios without zero component.

As shown in Fig. 9, three waveforms of duty ratios are produced by inverse Park's transformation. By applying these duty ratios (reference signal with fundamental frequency), it is expected that output voltage will have a frequency equal to 50 Hz. All these duty ratios are compared with a carrier signal (saw-toothed waveform) to generate the gate pulses. The frequency of the carrier is equal to the switching frequency $f_{sw} = 100$KHz. It is because of that the frequency of each generated gate pulse gets equal to switching frequency. These produced gate pulses are applied to the corresponding upper legs of the inverter (i.e., leg A, B, and C). For three lower legs, the produced gate pulses are reversed and then applied respectively. Figure 10 presents the carrier signal.

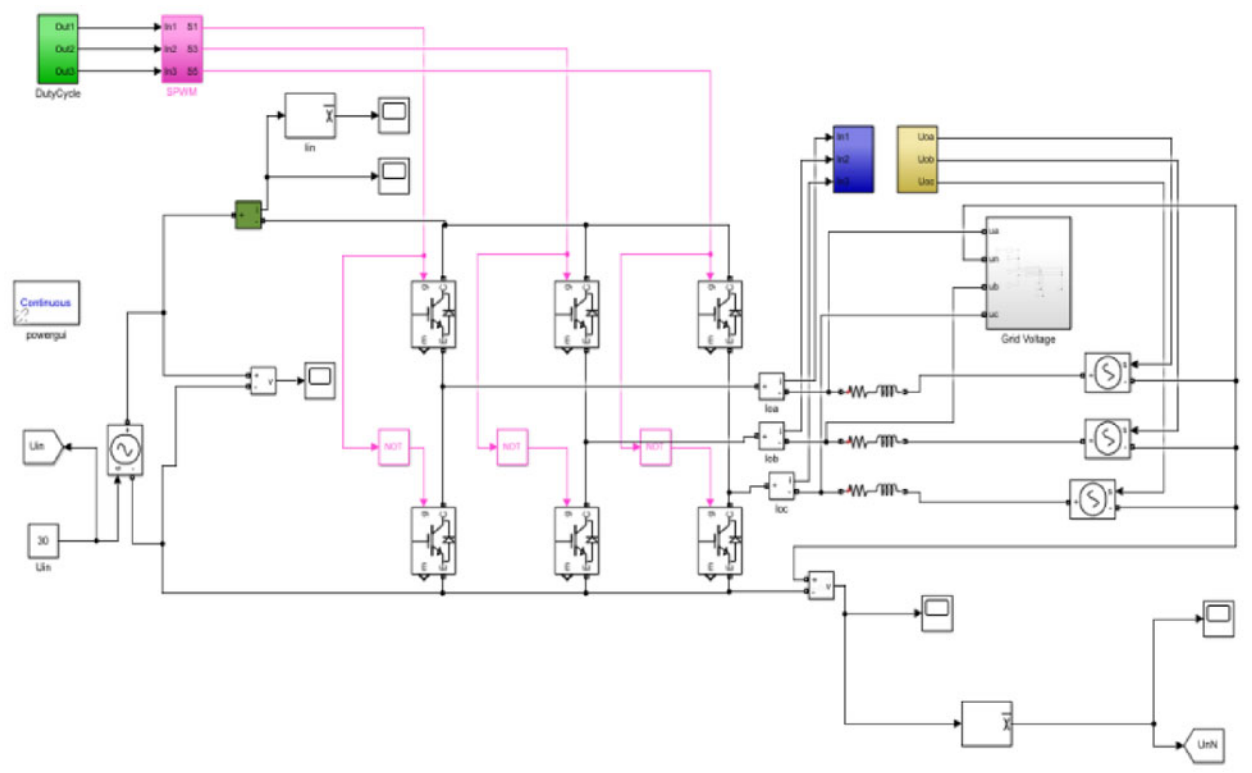

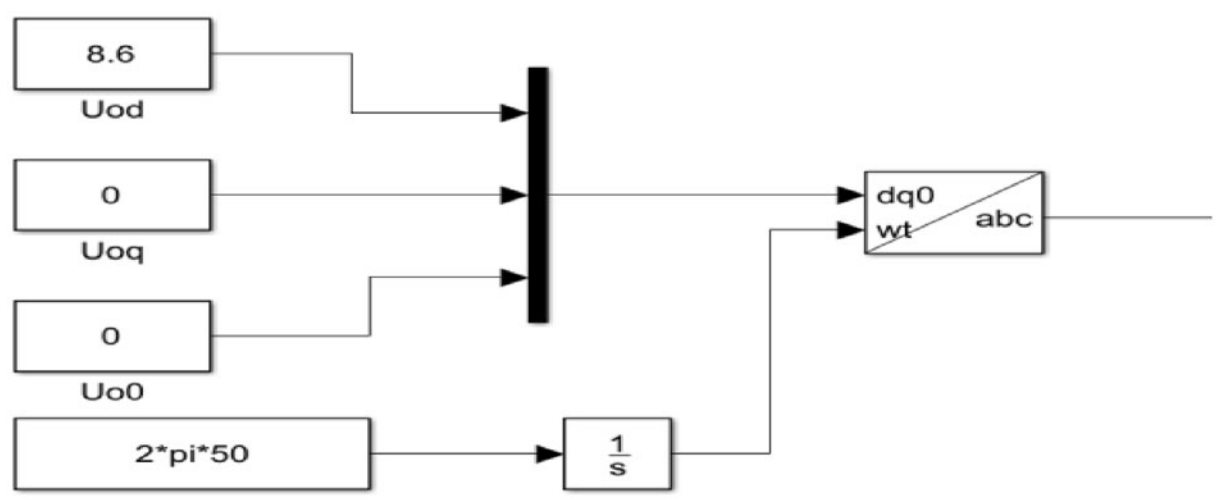

**Fig. 7** How to apply and generate the grid-connected output voltage of the converter

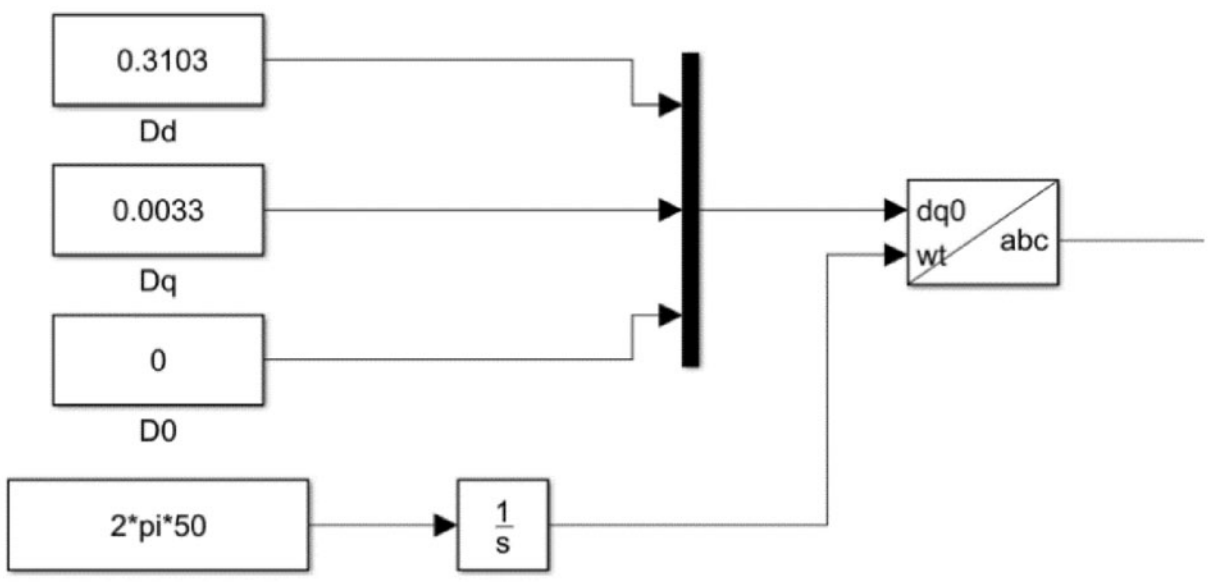

**Fig. 8** How to apply and generate duty ratio of the converter

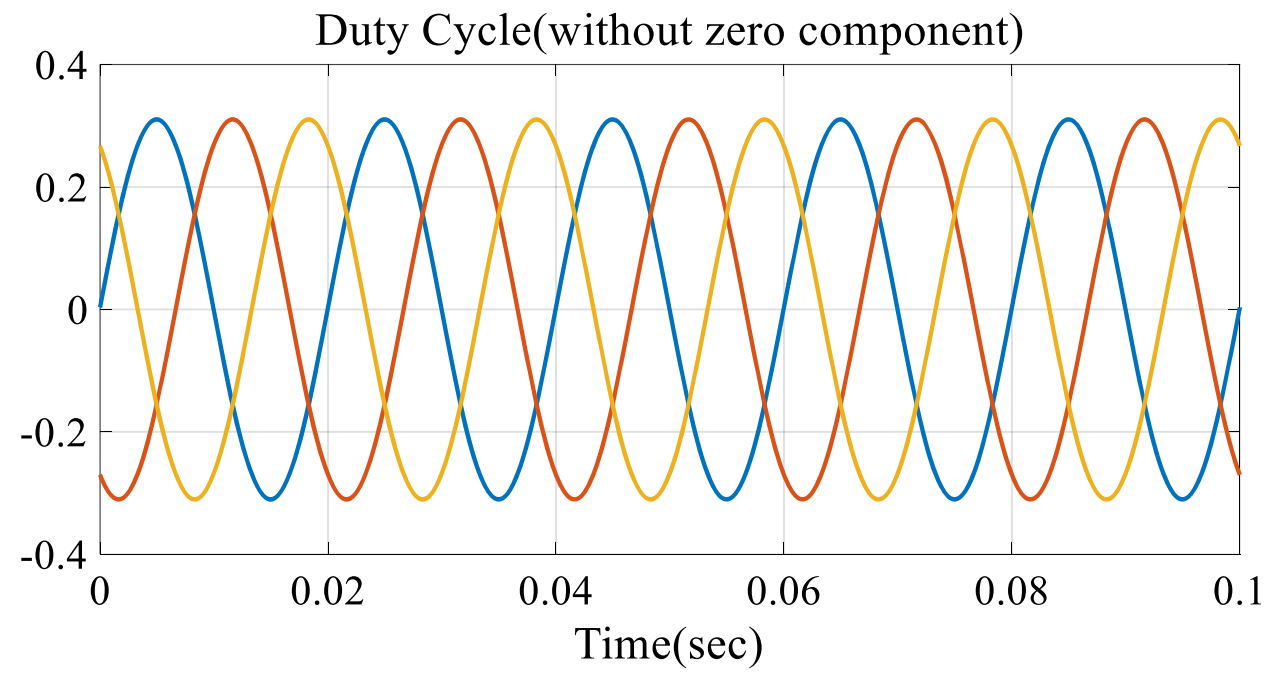

**Fig. 9** Waveforms of converter duty ratios in the absence of the zero component (synchronous frequency: 50 Hz)

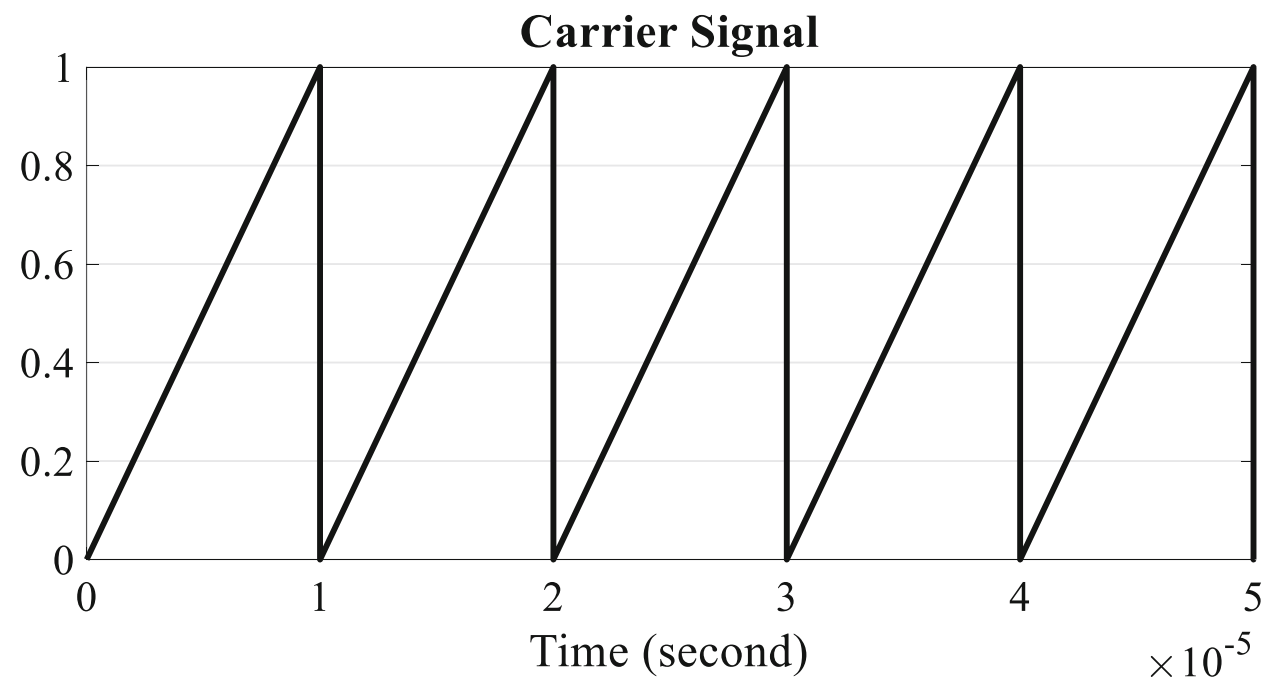

**Fig. 10** Carrier waveform with a frequency of 100 kHz (saw-toothed waveform)

However, as mentioned before and shown in Fig. 9, the duty ratios waveform has negative parts and they are symmetrical with respect to the time axis. Accordingly, there is a paradox here because the duty ratio is a value between 0 and 1 and cannot take negative values. The problem is that the zero component of the duty ratio has not yet been taken into account.

## 7 Problem formulation

The direct and inverse park's transformations are given as follows:

$$\begin{bmatrix} x_d \\ x_q \\ x_0 \end{bmatrix} = \overbrace{\frac{2}{3}\begin{bmatrix} \cos(\theta) & \cos\left(\theta - 2\frac{\pi}{3}\right) & \cos\left(\theta - 4\frac{\pi}{3}\right) \\ -\sin(\theta) & -\sin\left(\theta - 2\frac{\pi}{3}\right) & -\sin\left(\theta - 4\frac{\pi}{3}\right) \\ \frac{1}{2} & \frac{1}{2} & \frac{1}{2} \end{bmatrix}}^{T} \begin{bmatrix} x_a \\ x_b \\ x_c \end{bmatrix} \tag{75}$$

$$\begin{bmatrix} x_a \\ x_b \\ x_c \end{bmatrix} = \overbrace{\begin{bmatrix} \cos(\theta) & -\sin(\theta) & 1 \\ \cos\left(\theta - 2\frac{\pi}{3}\right) & -\sin\left(\theta - 2\frac{\pi}{3}\right) & 1 \\ \cos\left(\theta - 4\frac{\pi}{3}\right) & -\sin\left(\theta - 4\frac{\pi}{3}\right) & 1 \end{bmatrix}}^{T^{-1}} \begin{bmatrix} x_d \\ x_q \\ x_0 \end{bmatrix} \tag{76}$$

To obtain the zero component of duty ratio, the average model equations must be rewritten in matrix form as follows:

$$\frac{d}{dt}\begin{bmatrix} \langle i_{La} \rangle \\ \langle i_{Lb} \rangle \\ \langle i_{Lc} \rangle \end{bmatrix} = \frac{1}{L}\langle u_{in} \rangle \begin{bmatrix} d_A \\ d_B \\ d_C \end{bmatrix} - \frac{r_{eq}}{L}\begin{bmatrix} \langle i_{La} \rangle \\ \langle i_{Lb} \rangle \\ \langle i_{Lc} \rangle \end{bmatrix} - \frac{1}{L}\begin{bmatrix} \langle u_{an} \rangle \\ \langle u_{bn} \rangle \\ \langle u_{cn} \rangle \end{bmatrix} - \frac{\langle u_{nN} \rangle}{L}\begin{bmatrix} 1 \\ 1 \\ 1 \end{bmatrix} \tag{77}$$

In (77), the abc reference frame quantities can be replaced with their corresponding dq0 values in the a-b-c domain:

$$\frac{d}{dt}\left\{T^{-1}\begin{bmatrix}\langle i_{Ld}\rangle\\ \langle i_{Lq}\rangle\\ \langle i_{L0}\rangle\end{bmatrix}\right\} = \frac{1}{L}\langle u_{in}\rangle\left\{T^{-1}\begin{bmatrix}d_d\\ d_q\\ d_0\end{bmatrix}\right\} - \frac{r_{eq}}{L}\left\{T^{-1}\begin{bmatrix}\langle i_{Ld}\rangle\\ \langle i_{Lq}\rangle\\ \langle i_{L0}\rangle\end{bmatrix}\right\} - \frac{1}{L}\left\{T^{-1}\begin{bmatrix}\langle u_{od}\rangle\\ \langle u_{oq}\rangle\\ \langle u_{o0}\rangle\end{bmatrix}\right\} - \frac{\langle u_{nN}\rangle}{L}\begin{bmatrix}1\\1\\1\end{bmatrix} \tag{78}$$

By further simplification, we have:

$$\frac{d}{dt}\begin{bmatrix}\langle i_{Ld}\rangle\\ \langle i_{Lq}\rangle\\ \langle i_{L0}\rangle\end{bmatrix} = \frac{1}{L}\langle u_{in}\rangle\begin{bmatrix}d_d\\ d_q\\ d_0\end{bmatrix} - \overbrace{T\frac{d(T^{-1})}{dt}}^{\beta}\begin{bmatrix}\langle i_{Ld}\rangle\\ \langle i_{Lq}\rangle\\ \langle i_{L0}\rangle\end{bmatrix} - \frac{r_{eq}}{L}\begin{bmatrix}\langle i_{Ld}\rangle\\ \langle i_{Lq}\rangle\\ \langle i_{L0}\rangle\end{bmatrix} - \frac{1}{L}\begin{bmatrix}\langle u_{od}\rangle\\ \langle u_{oq}\rangle\\ \langle u_{o0}\rangle\end{bmatrix} - \frac{\langle u_{nN}\rangle}{L}\overbrace{T\begin{bmatrix}1\\1\\1\end{bmatrix}}^{\alpha} \tag{79}$$

For parts α and β, we have:

$$\alpha = \frac{2}{3}\begin{bmatrix}\cos(\theta) & \cos\left(\theta-2\frac{\pi}{3}\right) & \cos\left(\theta-4\frac{\pi}{3}\right)\\ -\sin(\theta) & -\sin\left(\theta-2\frac{\pi}{3}\right) & -\sin\left(\theta-4\frac{\pi}{3}\right)\\ \frac{1}{2} & \frac{1}{2} & \frac{1}{2}\end{bmatrix}\begin{bmatrix}1\\1\\1\end{bmatrix} = \frac{2}{3}\begin{bmatrix}0\\0\\ \frac{3}{2}\end{bmatrix} = \begin{bmatrix}0\\0\\1\end{bmatrix} \tag{80}$$

By considering $\theta = \omega t$ ($\omega$ being synchronous speed), β is obtained as (81):

$$\beta = \frac{2}{3}\begin{bmatrix}\cos(\omega t) & \cos\left(\omega t-2\frac{\pi}{3}\right) & \cos\left(\omega t-4\frac{\pi}{3}\right)\\ -\sin(\omega t) & -\sin\left(\omega t-2\frac{\pi}{3}\right) & -\sin\left(\omega t-4\frac{\pi}{3}\right)\\ \frac{1}{2} & \frac{1}{2} & \frac{1}{2}\end{bmatrix} \times \ldots \begin{bmatrix}-\omega\sin(\omega t) & -\omega\cos(\omega t) & 0\\ -\omega\sin\left(\omega t-2\frac{\pi}{3}\right) & -\omega\cos\left(\omega t-2\frac{\pi}{3}\right) & 0\\ -\omega\sin\left(\omega t-4\frac{\pi}{3}\right) & -\omega\cos\left(\omega t-4\frac{\pi}{3}\right) & 0\end{bmatrix} = \begin{bmatrix}0 & \omega & 0\\ -\omega & 0 & 0\\ 0 & 0 & 0\end{bmatrix} \tag{81}$$

Finally, Eq. (82) is obtained:

$$\frac{d}{dt}\begin{bmatrix}\langle i_{Ld}\rangle\\ \langle i_{Lq}\rangle\\ \langle i_{L0}\rangle\end{bmatrix} = \frac{1}{L}\langle u_{in}\rangle\begin{bmatrix}d_d\\ d_q\\ d_0\end{bmatrix} + \begin{bmatrix}-\frac{r_{eq}}{L} & \omega & 0\\ -\omega & -\frac{r_{eq}}{L} & 0\\ 0 & 0 & -\frac{r_{eq}}{L}\end{bmatrix}\begin{bmatrix}\langle i_{Ld}\rangle\\ \langle i_{Lq}\rangle\\ \langle i_{L0}\rangle\end{bmatrix} - \frac{1}{L}\begin{bmatrix}\langle u_{od}\rangle\\ \langle u_{oq}\rangle\\ \langle u_{o0}\rangle\end{bmatrix} - \frac{1}{L}\langle u_{nN}\rangle\begin{bmatrix}0\\0\\1\end{bmatrix} \tag{82}$$

The zero component of the duty ratio is obtained as follows:

$$\frac{d}{dt}\langle i_{L0}\rangle = \frac{1}{L}\left[-r_{eq}\langle i_{L0}\rangle - \langle u_{o0}\rangle - \langle u_{nN}\rangle + \langle u_{in}\rangle d_0\right] \tag{83}$$

By assuming that the grid voltages and currents are balanced, the left side of (83) can be equated zero. Consequently, the zero component of the duty ratio is obtained as follows:

$$d_0 = \frac{\langle u_{nN}\rangle}{u_{in}} \tag{84}$$

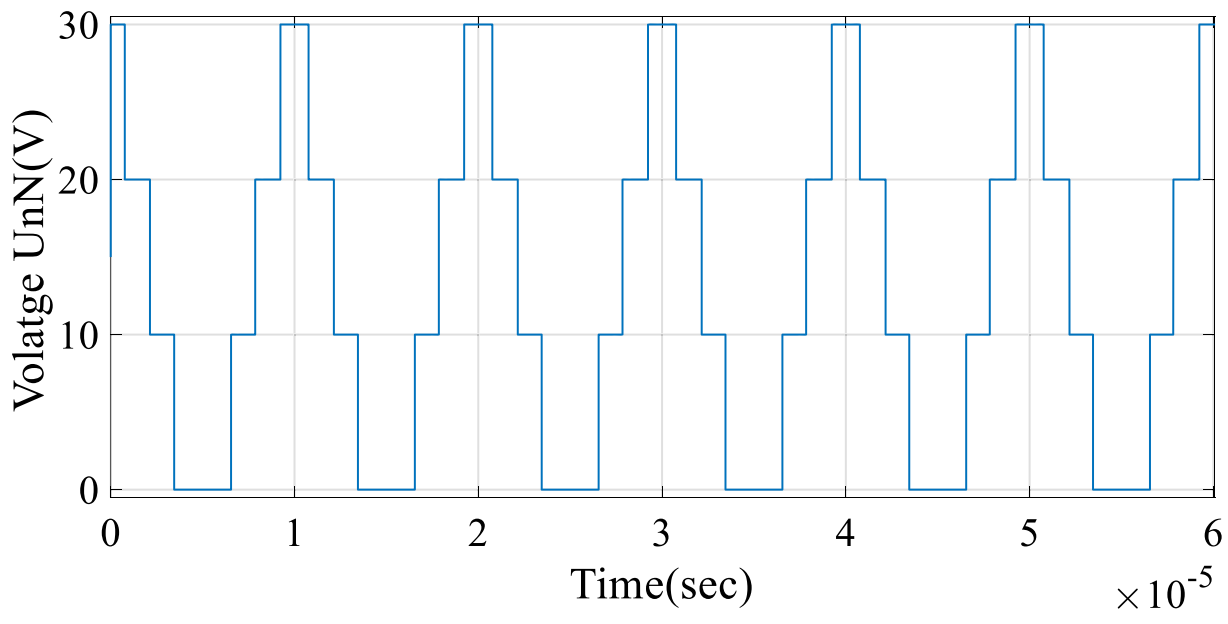

**Fig. 11** The simulation result of the instantaneous value of $u_{nN}$

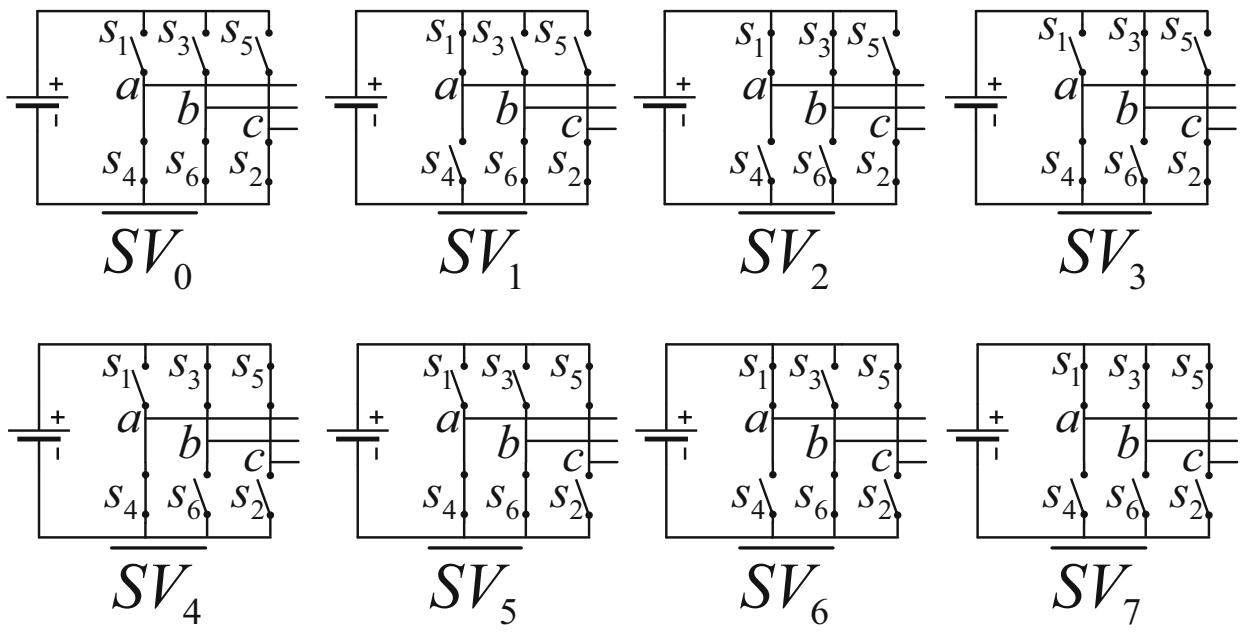

**Fig. 12** Eight possible switch combinations for VSI (Holmes and Lipo 2003)

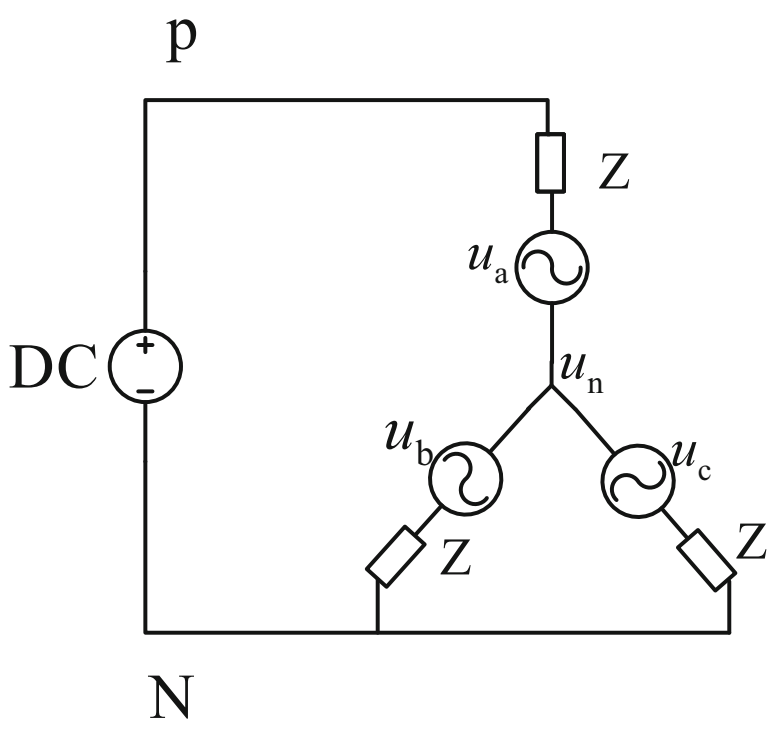

**Fig. 13** The equivalent circuit of the converter in the first active state ($SV_1$)

The numerator of Eq. (84), i.e.,$u_{nN}$, is the voltage of the grid neutral (n) with respect to the negative of the DC voltage source (N). Figure 11 presents the instantaneous values of $u_{nN}$:

In the space vector modulation (SVM) there are eight switch combinations (states) for a three-phase inverter. Two of these states (i.e.,$SV_0$ and $SV_7$) are called zero states or vectors because the output of the inverter is short-circuited. In the other six states, the input of the inverter is crossed to output (Holmes 2003). Figure 12 presents eight possible states of VSI.

As can be seen from Fig. 11, the instantaneous value of $u_{nN}$ varies periodically between zero and some positive value. This can be investigated by obtaining the equivalent circuit of the converter in different sectors. Here, the equivalent circuit of the converter is obtained for sectors 1 and 2 (active states), for example. The first equivalent circuit in sector one or active state $SV_1$ is as shown in Fig. 13.

Output current equations from $u_n$ are as follows:

$$i_a = \frac{u_n + u_a - u_N - U_{DC}}{Z} \tag{85}$$

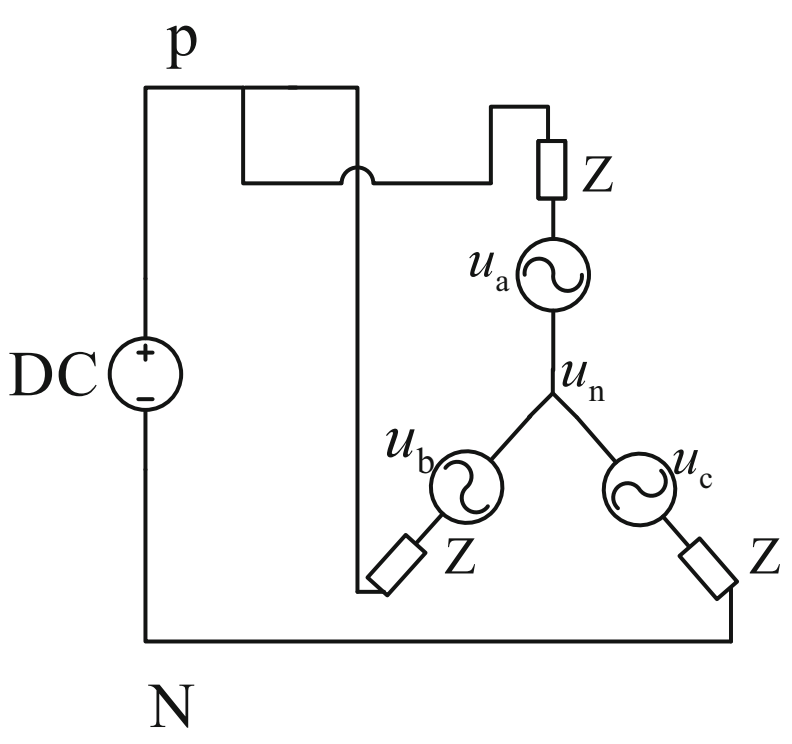

**Fig. 14** The equivalent circuit of the converter in the second active state ($SV_2$)

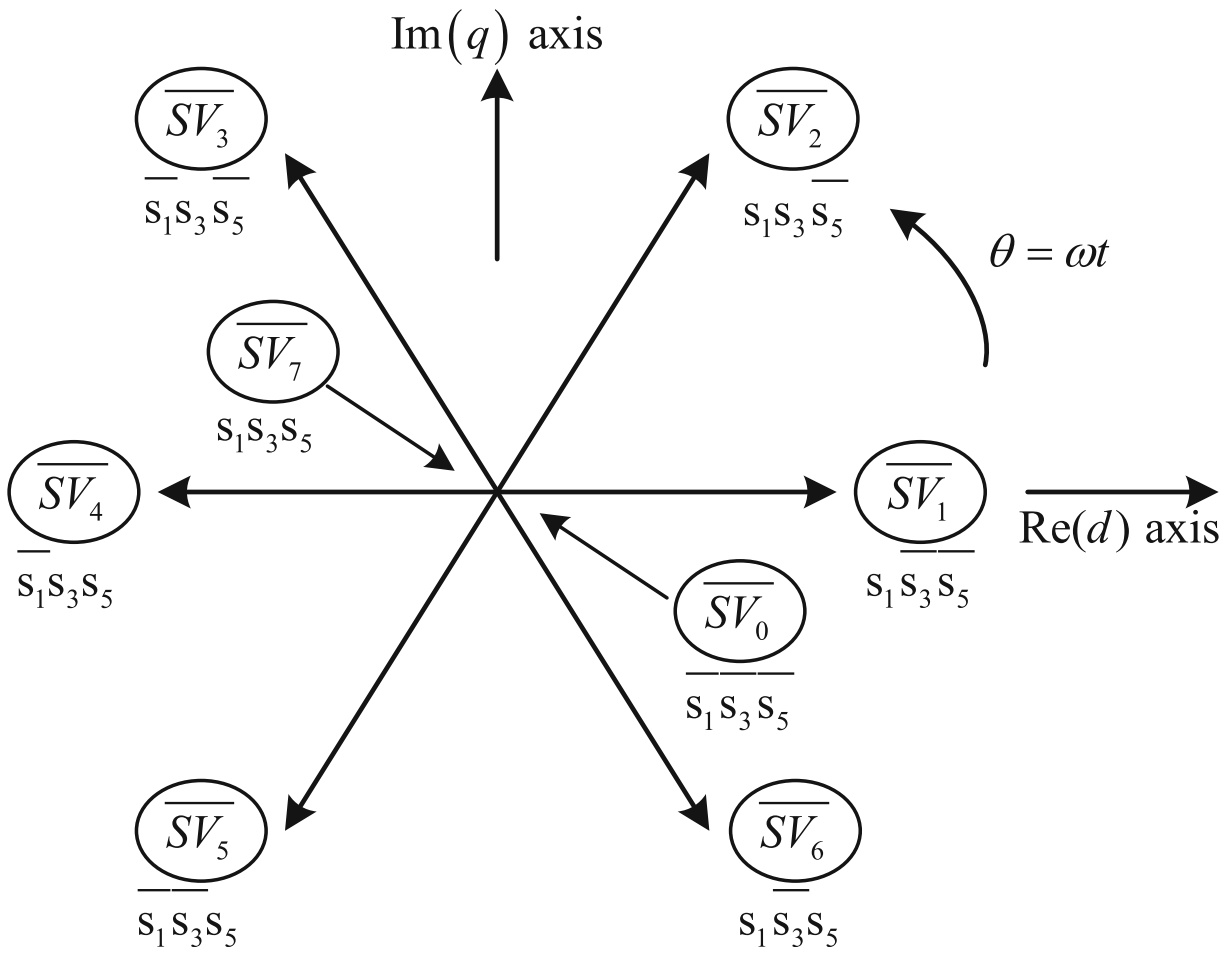

**Fig. 15** Eight possible stationary voltage vectors for a VSI (Holmes and Lipo 2003)

$$i_b = \frac{u_n + u_b - u_N}{Z} \tag{86}$$

$$i_c = \frac{u_n + u_c - u_N}{Z} \tag{87}$$

Assuming the balanced grid, these equations are expressed as follows:

$$i_a + i_b + i_c = \frac{3u_n - 3u_N - U_{DC} + \overbrace{(u_a + u_b + u_c)}^{=0}}{Z} = 0 \tag{88}$$

Therefore,$u_{nN}$ at the active state $SV_1$ is as below:

$$3u_{nN} - U_{DC} = 0$$
$$u_{nN} = \frac{U_{DC}}{3} \tag{89}$$

Figure 14 depicts the equivalent circuit of VSI for the active state ($SV_2$).

For this state, the current equations are as follows:

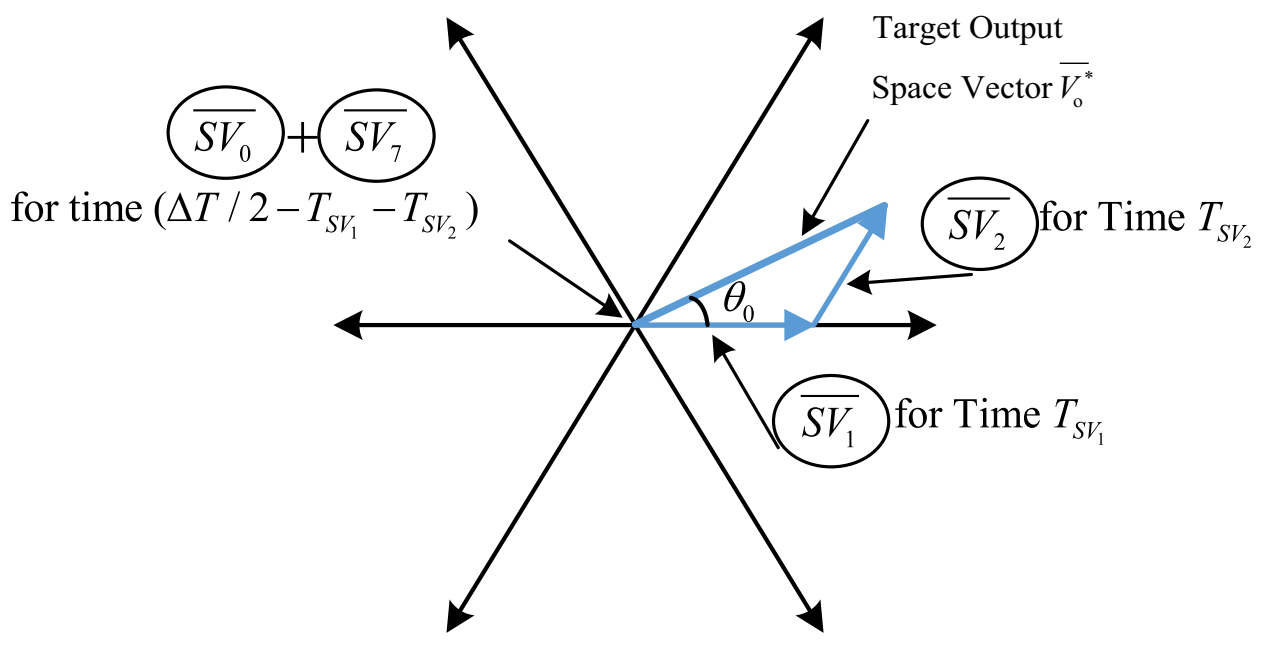

**Fig. 16** Construction of an arbitrary target voltage vector in sector one by geometrical summation of the two nearest space vectors (Holmes and Lipo 2003)

$$i_a = \frac{u_n + u_a - u_N - U_{DC}}{Z} \tag{90}$$

$$i_a = \frac{u_n + u_b - u_N - U_{DC}}{Z} \tag{91}$$

$$i_a = \frac{u_n + u_c - u_N}{Z} \tag{92}$$

And

$$i_a + i_b + i_c = \frac{3u_n - 3u_N - 2U_{DC} + \overbrace{(u_a + u_b + u_c)}^{=0}}{Z} = 0 \tag{93}$$

Therefore, the voltage $u_{nN}$ at the active state $SV_2$ is as follows:

$$u_{nN} = \frac{2U_{DC}}{3} \tag{94}$$

Figure 15 presents the locations of eight possible stationary voltage vectors for a VSI in the d-q reference frame (Holmes and Lipo 2003):

To implement SVM, one must first identify the sector where the arbitrary reference output voltage vector of the converter lies [11,19]. This vector (i.e.,$\overline{V_o^*}$) can be constructed by summing (averaging) a number of these space vectors ($\overline{SV}_{0\ldots7}$) within one switching cycle (period)$\Delta T/2$.

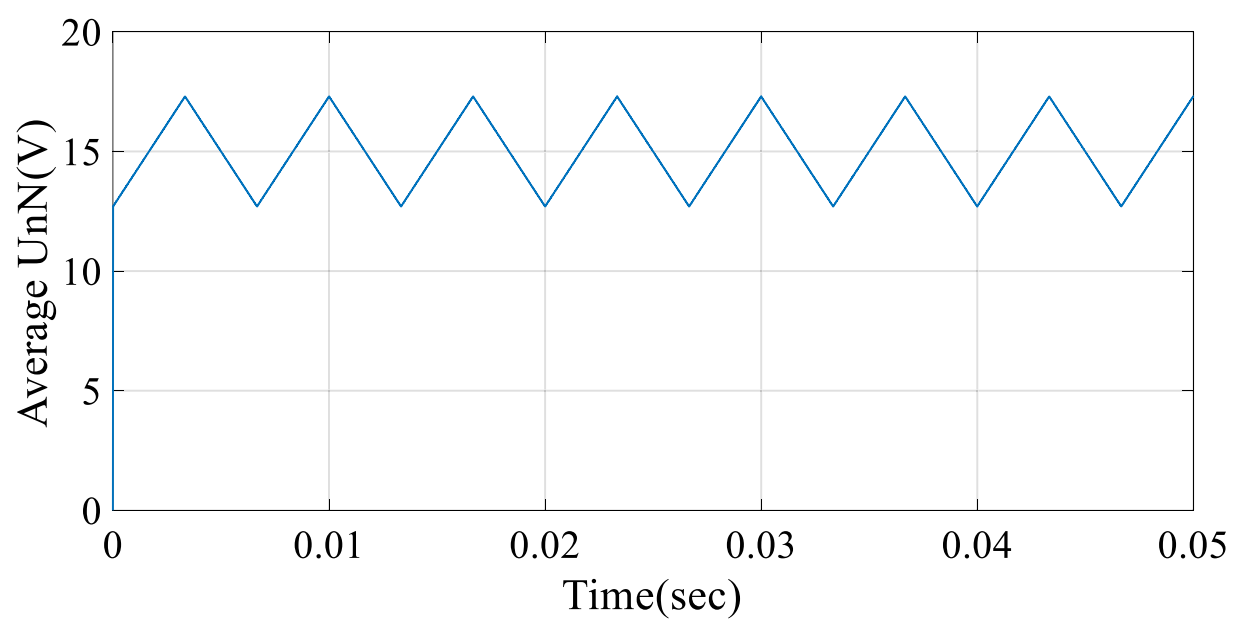

**Fig. 17** The simulation result for the average waveform of voltage $u_{nN}$ under the switching period

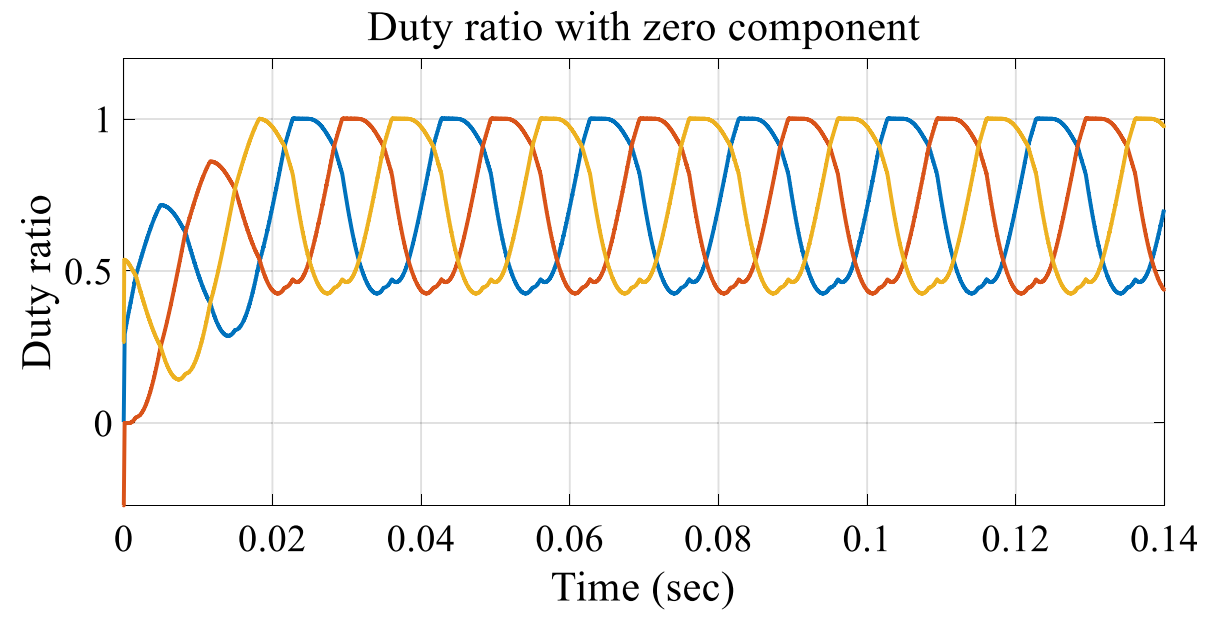

**Fig. 18** Waveforms of converter duty ratio with considering the zero component (synchronous frequency 50 Hz)

For example, see the geometry shown in Fig. 16 for the first 60° segment of the plane (Holmes and Lipo 2003):

To obtain the desired output vector $\overline{V_o^*}$, it is needed to know the active time of the space vectors ($\overline{SV}_{1\ldots6}$). In addition, it is necessary to know the sequence of space vectors ($\overline{SV}_{0\ldots7}$) that must be applied to create the output vector $\overline{V_o^*}$. The conventional SVM implementation puts the active space vectors in the center of each half switching period, and the remaining time is for the zero space vectors, which is split equally between $\overline{SV_0}$ and $\overline{SV_7}$. In the following, the space vector sequence is presented for sector 1 $(0 \le \theta_0 \le \pi/3)$ (Holmes and Lipo 2003):

$$\underbrace{\overline{SV_0} \to \overline{SV_1} \to \overline{SV_2} \to \overline{SV_7}}_{\Delta T/2} \to \underbrace{\overline{SV_7} \to \overline{SV_2} \to \overline{SV_1} \to \overline{SV_0}}_{\Delta T/2} \tag{95}$$

Here, it is desired to obtain the voltage $u_{nN}$ to find the zero component of the duty ratio. So, by (94) in sector 1, the $u_{nN}$ will be as follows:

The values and sequence of voltage $u_{nN}$ for sector 2 are as follows:

The values of these tables and analytical results are in agreement with the waveform obtained by simulation in Fig. 11. Figure 17 shows the average waveform of voltage $u_{nN}$ under the switching period obtained by the simulation (Tables 3, 4):

As expected, the average waveform of voltage $u_{nN}$ is non-zero and positive.

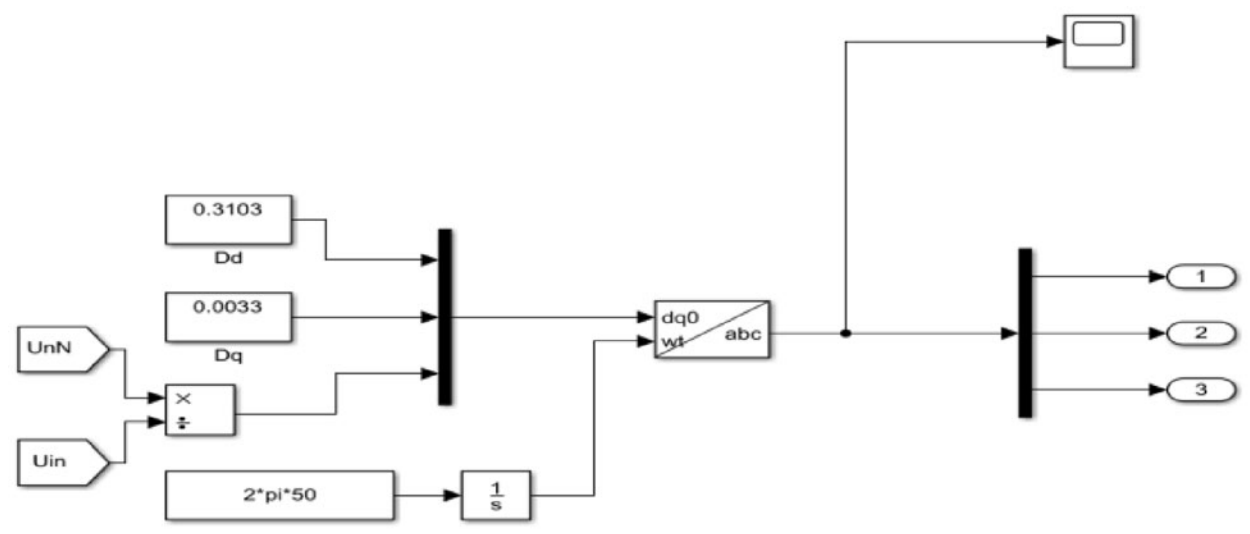

**Fig. 19** The inverse park's transformation to produce a duty ratio in the a-b-c domain in the Simulink

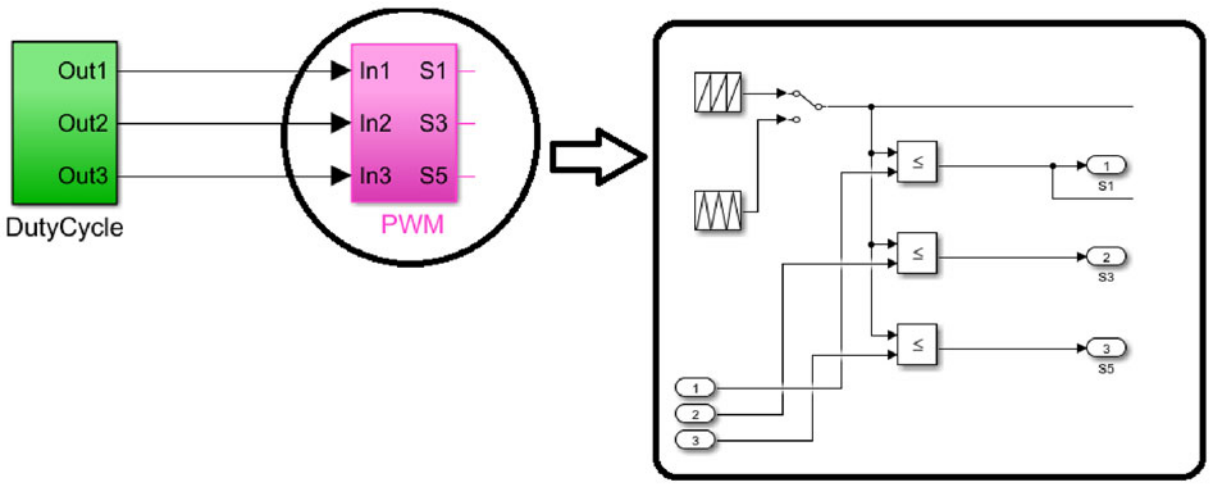

**Fig. 20** Setup for producing the gate pulses

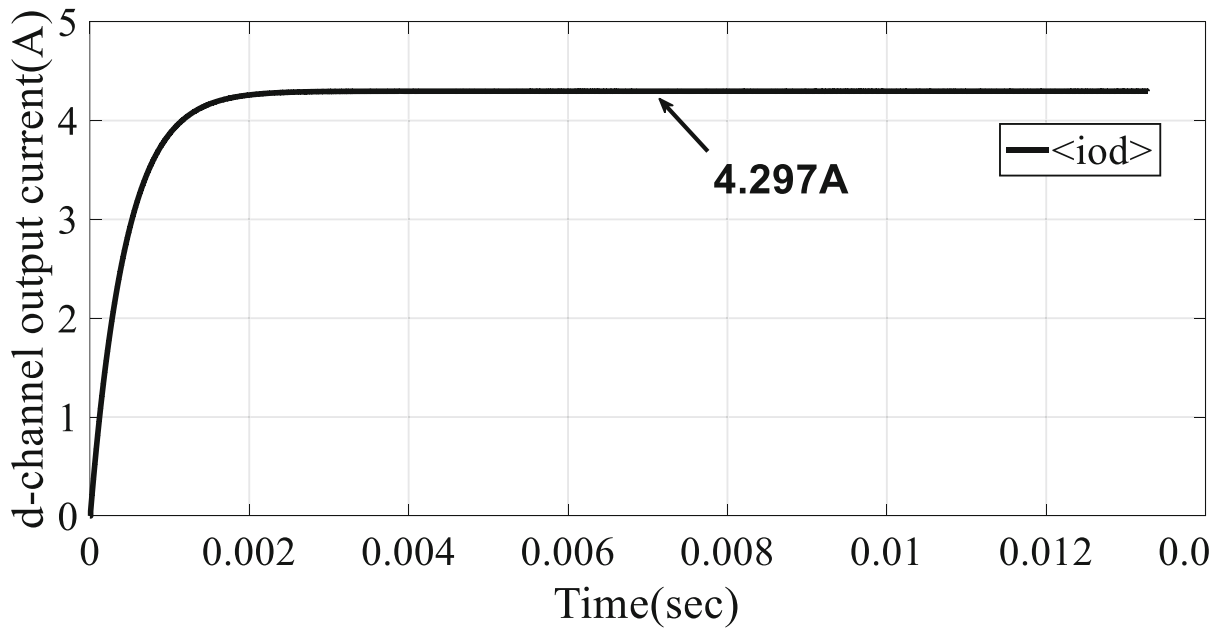

**Fig. 21** The average waveform of d-channel output current $\langle i_{od} \rangle$

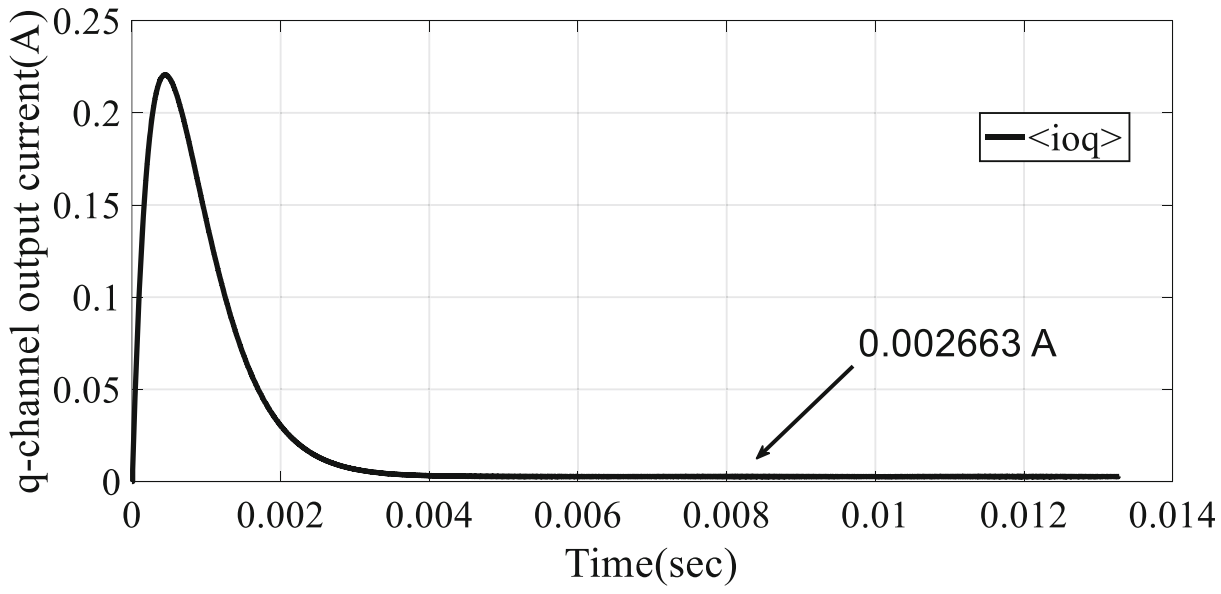

**Fig. 22** The average waveform of q-channel output current $\langle i_{oq} \rangle$

Figure 18 shows the waveforms of the duty ratios obtained in the a-b-c domain where the zero sequence of the duty ratio is considered.

As can be seen, the waveforms of duty ratios are all placed above the time axis and take positive values. In fact, these waveforms do not contradict the duty ratio definition, which must have values between 0 and 1. The zero components of grid-connected voltage and grid current are zero because they are assumed balanced while the duty ratio has a zero component. Figure 19 illustrates the simulation setup to obtain these waveforms is depicted.

To generate the gate pulses of three upper IGBT's of the inverter, the setup in Fig. 20 is used.

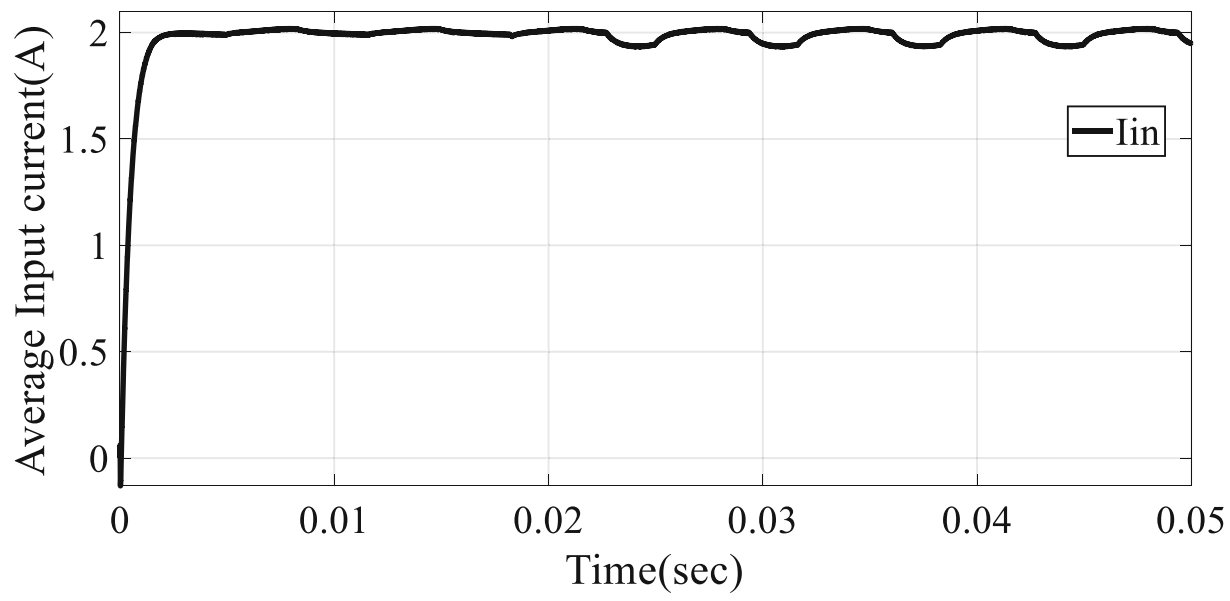

**Fig. 23** The average waveform of input current $\langle i_{in} \rangle$

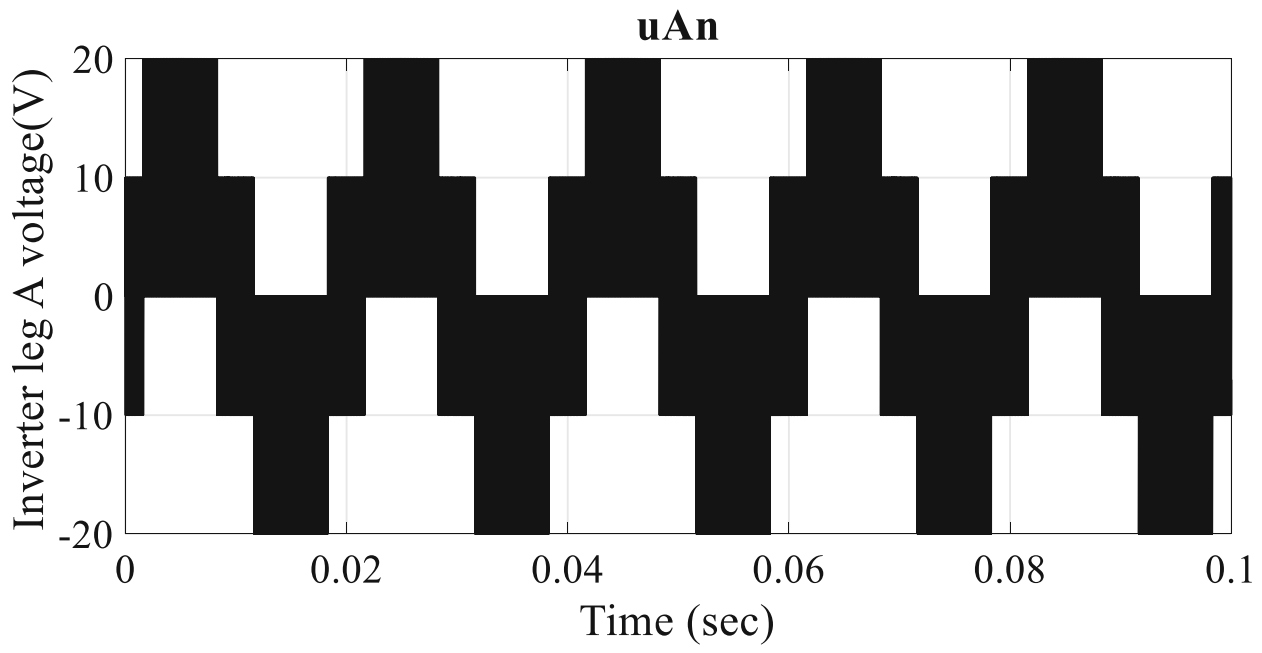

**Fig. 24** The waveform of the output voltage of leg A

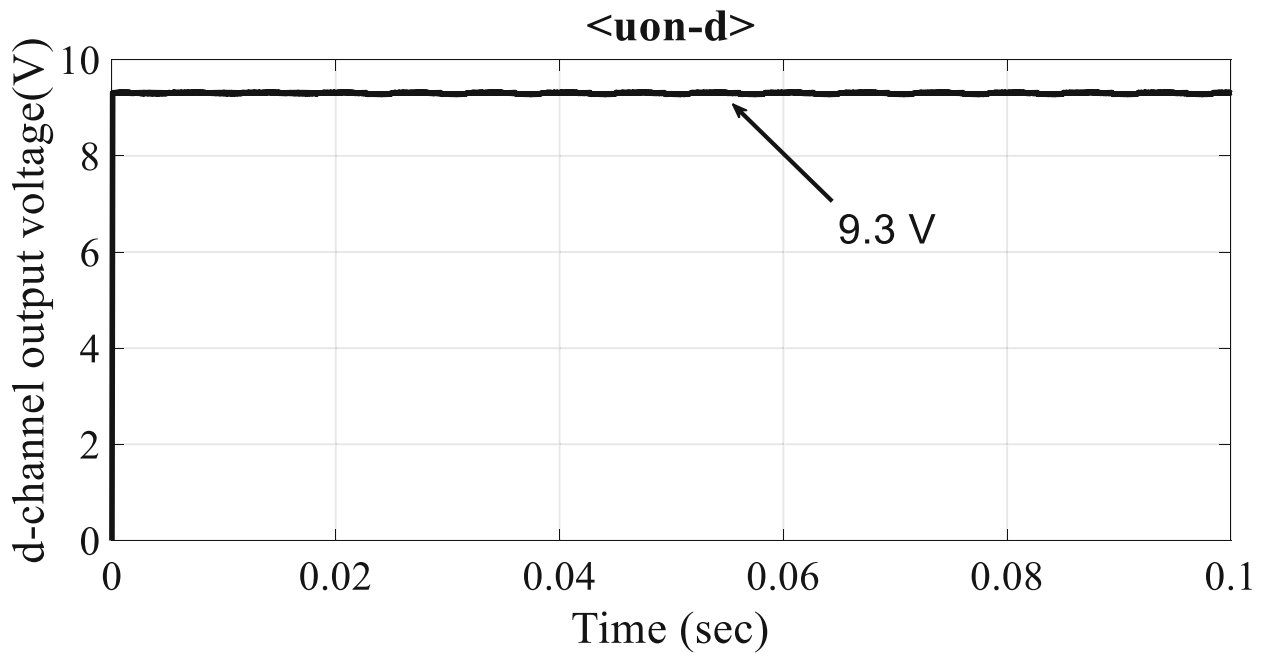

**Fig. 25** The d-channel of output

## 8 Simulation results

In this section, the waveforms of switching time averaged d-q components of inverter output currents $\langle i_{od} \rangle, \langle i_{oq} \rangle$, obtained analytically in Sect. 3, are verified by time domain simulations using Matlab® Simulink. It is noteworthy that from Fourier analysis, the DC components of output currents $i_{od}, i_{oq}$ are given by their average values (Erickson and Dragan 2007). Figure 21 shows the waveform of the d-channel average output current $\langle i_{od} \rangle$.

As can be seen, the final value of $\langle i_{od} \rangle$ is approximately equal to the DC component obtained analytically, that is $I_{od} = 4.4236$. For the q-channel average output current, the waveform will be as shown in Fig. 22.

The final value of $\langle i_{oq} \rangle$ is nearly zero, which is equal to the DC component; i.e.,$I_{oq} = 0$. In the simulations, the circuit has been assumed to be at rest initially; that is all the inductor currents are zero at $t = 0$. The natural (ABC) domain currents being zero, the d-axis and q-axis current components will also be initially zero. This can be verified in the body of the paper by viewing Figs. 21 and 22, showing the switching averaged d-axis and q-axis currents, respectively, where it is clear that both the currents start at zero values. Considering the power control strategy, the reactive power is controlled to be zero (i.e. unity power factor operation). This strategy necessitates having zero q-axis current. Therefore, as visible in Fig. 22, the final value of the q-axis current is also zero, like its initial value. Between two steady state operating points (i.e. one at $t = 0$ and the other at $t \to \infty$) there will be current transients. This is why the q-axis current has a so called overshoot in Fig. 22.

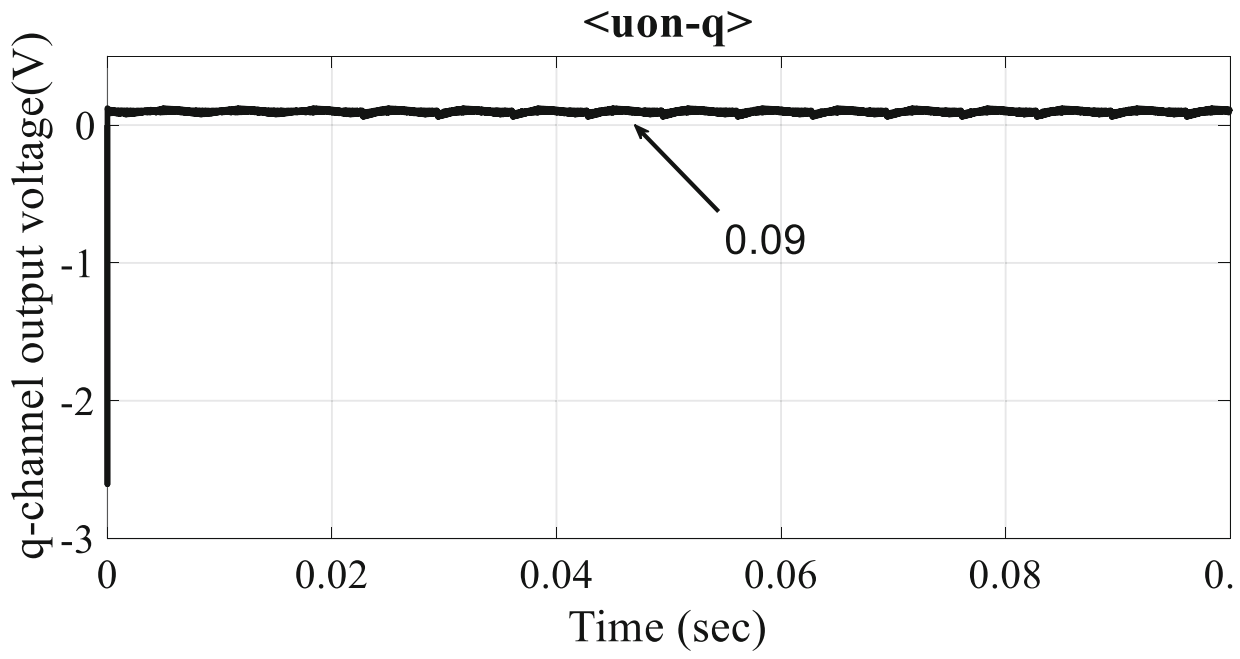

**Fig. 26** The q-channel of output

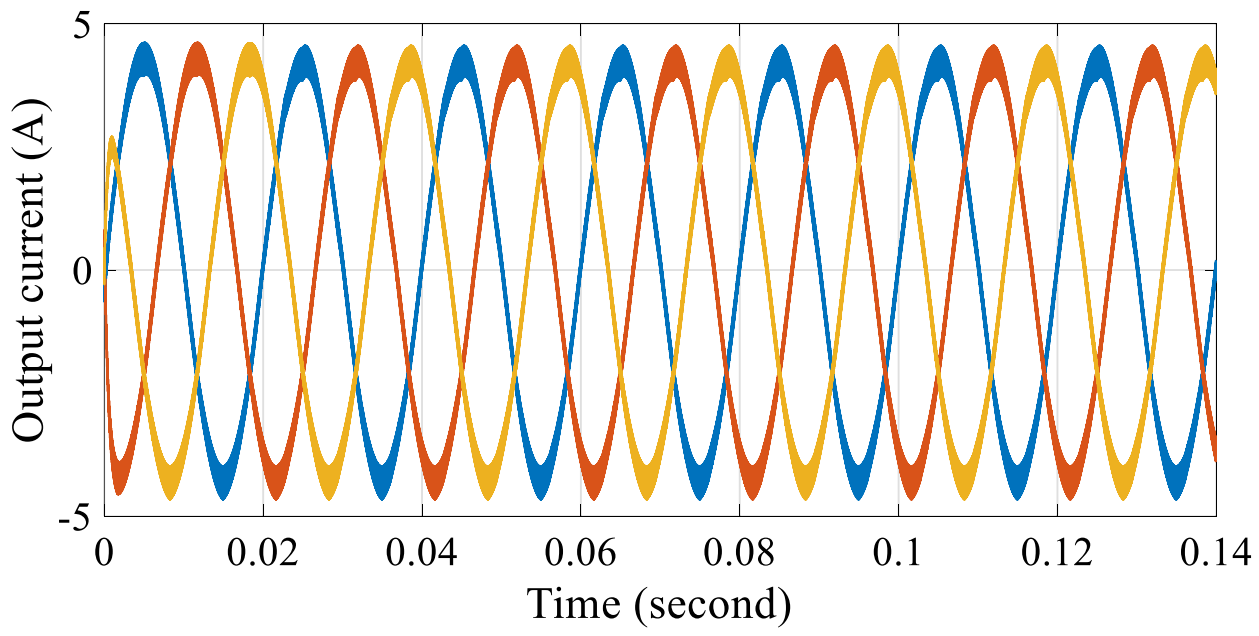

**Fig. 27** Output currents

**Table 3** The voltage $u_{nN}$ in sector 1

| Sector1 | $\overline{SV_0}$ | $\overline{SV_1}$ | $\overline{SV_2}$ | $\overline{SV_7}$ | $\overline{SV_7}$ | $\overline{SV_2}$ | $\overline{SV_1}$ | $\overline{SV_0}$ |
|---|---|---|---|---|---|---|---|---|
| $u_{nN}$ | 0 | 10 | 20 | 30 | 30 | 20 | 10 | 0 |

**Table 4** The voltage $u_{nN}$ in sector2

| Sector1 | $\overline{SV_0}$ | $\overline{SV_3}$ | $\overline{SV_2}$ | $\overline{SV_7}$ | $\overline{SV_7}$ | $\overline{SV_2}$ | $\overline{SV_3}$ | $\overline{SV_0}$ |
|---|---|---|---|---|---|---|---|---|
| $u_{nN}$ | 0 | 10 | 20 | 30 | 30 | 20 | 10 | 0 |

The switching time averaged waveform of the input current is demonstrated in Fig. 23.

As can be seen, the simulated result is approximately identical with the predefined value, which is $I_{in} = 2$A.

Figure 24 presents the output voltage of leg A of the inverter relative to the neutral of the star point $u_{An}$. As can be seen from Fig. 24, the VF-VSI has buck-type characteristics (puukko 2012). The peak value of output voltage is $u_{(A,B,C)} = 2U_{in}/3$.

The dq component of the output voltage is as given in Fig. 25.

And, for its q component, the waveform is given in Fig. 26.

Finally, Fig. 27 presents the three-phase output currents.

## 9 Conclusion

In this paper, a detailed overview of the dynamic modeling of the grid-connected voltage fed inverter is performed and the large-signal and small-signal converter equations are obtained. It was explained that the small-signal model can be achieved by linearizing averaged large-signal equations around a quiescent operating point. The reason for the averaging of converter equations is that one can eliminate the switching ripple and switching harmonics by averaging over a switching period from the converter inductor voltage and capacitor current. In fact, by averaging the converter equations, we obtain the large-signal model of the inverter. In general, large-signal equations involve a set of nonlinear terms. The waveforms of the converter also contain harmonics of the modulation frequency. To obtain a linear model that is easier to analyze, it is essential to construct a small-signal model linearized around an operating point in such a manner that neglects the harmonics of the modulation. To accurately predict the poles and zeros of the inverter small-signal transfer function, we must linearize the large-signal averaged equations. The difference in modeling of DC-DC converters and DC-AC converters is that unlike the DC-DC converters, even in the steady-state or in equilibrium the net change in inductor current and capacitor voltage over one switching period is non-zero. This result is attributed to the nature of the AC values that are not constant even at steady state. Transforming AC electrical quantities into constant DC quantities by Park's transformation is common for analyzing steady-state stability of the power electronic converters. To apply the SSA method to the inverter, we need to have constant values at steady state.

To verify the model results, we apply the duty cycles obtained from our analysis to the inverter in Matlab® Simulink and compare the obtained values and waveforms with those from Sect. 3. The difference of this paper with other works is that the zero component of the duty ratio is taken into account. This means that to apply the obtained equilibrium duty ratios in the d-q domain to the inverter in Simulink®, we consider the zero component as well as the d and q components in inverse park's transformation for producing the duty ratios in the a-b-c domain. Actualy, if this is not taken into account, the duty ratios achieved in the a-b-c domain contradict the definition of duty ratio, which must be between 0 and 1. This is done by rewriting the averaged equations of the converter into matrix form.

The method is proved to be correct in general; numerical simulations in the paper are just to make a verification of the theory behind the subject for a particular case.

In this study, we provide a tool to obtain and verify the converter transfer function for stability analysis and controller design in the d-q domain for the converter. In future works, these transfer functions can be verified by Matlab® Simulink toolboxes. Afterward, the controller based on various transfer functions can be designed and optimized.

**Declarations**

**Conflict of interest** The authors have no conflicts of interest to declare that are relevant to the content of this article.